\documentclass[12pt,preprint,flushrt]{aastex}
\slugcomment{\it Accepted to The Astrophysical Journal (Nov. 26, 2001)} 
\lefthead{Stern et al.}
\righthead{Type II Quasar}

\def\ie{{i.e.,~}}
\def\eg{{e.g.,~}}
\def\etal{{et al.~}}

\def\deg{\ifmmode {^{\circ}}\else {$^\circ$}\fi}

\def\secper{\ifmmode \rlap.{^{s}}\else $\rlap{.}{^{s}} $\fi}

\def\kms{\ifmmode {\rm\,km\,s^{-1}}\else
    ${\rm\,km\,s^{-1}}$\fi}
\def\kmsMpc{\ifmmode {\rm\,km\,s^{-1}\,Mpc^{-1}}\else
    ${\rm\,km\,s^{-1}\,Mpc^{-1}}$\fi}
\def\ergAcm2{\ifmmode {\rm\,ergs\,cm^{-2}\,{\rm \AA}^{-1}}\else
    ${\rm\,ergs\,cm^{-2}\,\AA^{-1}}$\fi}
\def\cm2{\ifmmode {\rm\,cm^{-2}}\else
    ${\rm\,cm^{-2}}$\fi}
\def\ergcm2s{\ifmmode {\rm\,ergs\,cm^{-2}\,s^{-1}}\else
    ${\rm\,ergs\,cm^{-2}\,s^{-1}}$\fi}
\def\cgsdeg2{\ifmmode {\rm\,ergs\,cm^{-2}\,s^{-1}\,deg^{-2}}\else
    ${\rm\,ergs\,cm^{-2}\,s^{-1}\,deg^{-2}}$\fi}
\def\ergsHz{\ifmmode {\rm\,ergs\,s^{-1}\,Hz^{-1}}\else
    ${\rm\,ergs\,s^{-1}\,Hz^{-1}}$\fi}
\def\ergs{\ifmmode {\rm\,ergs\,s^{-1}}\else
    ${\rm\,ergs\,s^{-1}}$\fi}
\def\ergsA{\ifmmode {\rm\,ergs\,s^{-1}\,\AA^{-1}}\else
    ${\rm\,ergs\,s^{-1}\,\AA^{-1}}$\fi}
\def\WHz{\ifmmode {\rm\,W\,Hz^{-1}}\else
    ${\rm\,W\,Hz^{-1}}$\fi}
\def\WHzsr{\ifmmode {\rm\,W\,Hz^{-1}\,sr^{-1}}\else
    ${\rm\,W\,Hz^{-1}\,sr^{-1}}$\fi}
\def\ergscm2Hz{\ifmmode {\rm\,ergs\,cm^{-2}\,s^{-1}\,Hz^{-1}}\else
    ${\rm\,ergs\,cm^{-2}\,s^{-1}\,Hz^{-1}}$\fi}

\def\spose#1{\hbox to 0pt{#1\hss}}
\def\simlt{\mathrel{\spose{\lower 3pt\hbox{$\mathchar"218$}}
     \raise 2.0pt\hbox{$\mathchar"13C$}}}
\def\simgt{\mathrel{\spose{\lower 3pt\hbox{$\mathchar"218$}}
     \raise 2.0pt\hbox{$\mathchar"13E$}}}

\def\ovi{\ion{O}{6} $\lambda$1035}
\def\lya{Ly$\alpha$}
\def\nv{\ion{N}{5} $\lambda$1240}

\def\sioiv{\ion{Si}{4}/\ion{O}{4}] $\lambda$1403}
\def\civ{\ion{C}{4} $\lambda$1549}
\def\heii{\ion{He}{2} $\lambda$1640}
\def\oiiip{\ion{O}{3}] $\lambda$1663}

\def\ciii{\ion{C}{3}] $\lambda$1909}

\def\oii{[\ion{O}{2}] $\lambda$3727}
\def\oiipair{[\ion{O}{2}] $\lambda \lambda$3726,3729}

\def\oiiia{[\ion{O}{3}] $\lambda$4959}
\def\oiii{[\ion{O}{3}] $\lambda$5007}

\def\nii{[\ion{N}{2}] $\lambda$6584}
\def\sii{[\ion{S}{2}] $\lambda \lambda$6716,6731}

\def\plotfiddle#1#2#3#4#5#6#7{\centering \leavevmode
\vbox to#2{\rule{0pt}{#2}}
\includegraphics{#1}}

\begin{document}

\title{{\it Chandra} Detection of a Type~II Quasar at $z=3.288^1$}

\author{Daniel Stern\altaffilmark{2}, Edward C. Moran\altaffilmark{3},
Alison L. Coil\altaffilmark{3}, Andrew Connolly\altaffilmark{4},
Marc Davis\altaffilmark{3,5}, \\ Steve Dawson\altaffilmark{3}, 
Arjun Dey\altaffilmark{6}, Peter Eisenhardt\altaffilmark{2}, 
Richard Elston\altaffilmark{7}, James R. Graham\altaffilmark{3}, \\
Fiona Harrison\altaffilmark{8}, David J. Helfand\altaffilmark{9},
Brad Holden\altaffilmark{10,11}, Peter Mao\altaffilmark{8}, 
Piero Rosati\altaffilmark{12}, Hyron Spinrad\altaffilmark{3}, 
S.A. Stanford\altaffilmark{10,11}, Paolo Tozzi\altaffilmark{13,14},
\& K.L. Wu\altaffilmark{7}}

\altaffiltext{1}{Some of the data presented herein were obtained at the
W.M. Keck Observatory, which is operated as a scientific partnership
among the California Institute of Technology, the University of
California and the National Aeronautics and Space Administration. The
Observatory was made possible by the generous financial support of the
W.M. Keck Foundation.  Based on observations with the NASA/ESA {\it
Hubble Space Telescope}, obtained at the Space Telescope Science
Institute, which is operated by the Association of Universities for
Research in Astronomy, Inc. under NASA contract No. NAS5-26555.  Based
on observations at the Kitt Peak National Observatory, National Optical
Astronomy Observatory, which is operated by the Association of
Universities for Research in Astronomy, Inc. under cooperative
agreement with the National Science Foundation.}

\altaffiltext{2}{Jet Propulsion Laboratory, California Institute of
Technology, Mail Stop 169-327, Pasadena, CA 91109 USA [{\tt email:
stern@zwolfkinder.jpl.nasa.gov}]}

\altaffiltext{3}{Department of Astronomy, University of California at
Berkeley, Berkeley, CA 94720 USA}

\altaffiltext{4}{Department of Physics and Astronomy, University of
Pittsburgh, Pittsburgh, PA 15260 USA}

\altaffiltext{5}{Department of Physics, University of California at
Berkeley, Berkeley, CA 94720 USA}

\altaffiltext{6}{Kitt Peak National Observatory, 950 North Cherry
Avenue, Tucson, AZ 85719 USA}

\altaffiltext{7}{Department of Astronomy, The University of Florida,
P.O. Box 112055, Gainesville, FL 32611 USA}

\altaffiltext{8}{Division of Physics, Mathematics and Astronomy,
105$-$24, California Institute of Technology, Pasadena, CA 91125 USA}

\altaffiltext{9}{Columbia University, Department of Astronomy, 550 West
120th Street, New York, NY 10027 USA}

\altaffiltext{10}{Institute of Geophysics and Planetary Physics,
Lawrence Livermore National Laboratory, L-413, Livermore, CA 94550 USA}

\altaffiltext{11}{Physics Department, University of California at Davis,
Davis, CA 95616 USA}

\altaffiltext{12}{European Southern Observatory, Karl-Schwarzschildstr.
2, D-85748, Garching, Germany}

\altaffiltext{13}{Department of Physics and Astronomy, The Johns Hopkins
University, Baltimore, MD 21218 USA}

\altaffiltext{14}{Osservatorio Astronomico di Trieste, via G.B. Tiepolo
11, I-34131, Trieste, Italy}

\begin{abstract}

We report on observations of a Type~II quasar at redshift $z = 3.288$,
identified as a hard X-ray source in a 185~ks observation with the {\it
Chandra X-ray Observatory} and as a high-redshift photometric candidate
from deep, multiband optical imaging.  CXO~J084837.9+445352
(hereinafter CXO52) shows an unusually hard X-ray spectrum from which
we infer an absorbing column density $N_{\rm H} = (4.8 \pm 2.1) \times
10^{23}$ cm$^{-2}$ (90\% confidence) and an implied unabsorbed 2$-$10
keV rest-frame luminosity of $L_{2-10} = 3.3 \times 10^{44}$ \ergs,
well within the quasar regime.  {\it Hubble Space Telescope} imaging
shows CXO52 to be elongated with slight morphological differences
between the WFPC2 F814W and NICMOS F160W bands.  Optical and
near-infrared spectroscopy of CXO52 show high-ionization emission lines
with velocity widths $\sim 1000\ \kms$ and flux ratios similar to a
Seyfert~2 galaxy or radio galaxy.  The latter are the only class of
high-redshift Type~II luminous AGN which have been extensively studied
to date.  Unlike radio galaxies, however, CXO52 is radio quiet,
remaining undetected at radio wavelengths to fairly deep limits,
$f_{\rm 4.8GHz} < 40 \mu$Jy.  High-redshift Type~II quasars, expected
from unification models of active galaxies and long-thought necessary
to explain the X-ray background, are poorly constrained observationally
with few such systems known.  We discuss recent observations of similar
Type~II quasars and detail search techniques for such systems:  namely
(1) X-ray selection, (2) radio selection, (3) multi-color imaging
selection, and (4) narrow-band imaging selection.  Such studies are
likely to begin identifying luminous, high-redshift Type~II systems in
large numbers.  We discuss the prospects for these studies and their
implications to our understanding of the X-ray background.

\end{abstract}


\section{Introduction}

Unified models of active galactic nuclei (AGN) posit orientation and
intrinsic luminosity as the two primary physical parameters governing
the optical characteristics of an active galaxy \markcite{Antonucci:93,
Urry:95}(\eg Antonucci 1993; Urry \& Padovani 1995).  According to
these models, when the obscuring torus of an AGN is oriented
perpendicular to the observer line of sight, one sees
spatially-unresolved, Doppler-boosted emission from hot gas near the
central supermassive black hole in the form of broad (FWHM $= 5000 -
20000 \kms$), high-ionization emission lines and synchrotron emission
from the central nucleus in the form of power-law ultraviolet/optical
continuum emission.  Independent of orientation one sees narrow
emission lines (FWHM $=500-1500 \kms$) from the spatially-extended
narrow-line region.  When the line of sight to the central engine is
blocked by the torus, only the extended narrow-line emission is seen.
Active galaxies without observed broad emission lines but with
high-ionization species are classified as Type~II systems, while those
with broad line emission are classified Type~I.  In addition to the
narrow line region, hard X-ray, far-infrared, and radio emission from
AGN are also largely immune to obscuration and orientation effects.

Intrinsic, unobscured luminosity, or activity level, is the second
primary physical parameter differentiating AGN.  The most luminous
active galaxies ($-30 \simlt M_{\rm B} \simlt -23$), where light from
the central regions overwhelm evidence of the underlying stellar
populations, are identified as quasars (QSOs).  At lower luminosity
($M_B \simgt -23$), Seyfert galaxies show evidence of both host galaxy
stellar emission and radiation from the central AGN.  Finally, surveys
of local galaxies find a class of sources with very low power,
low-ionization nuclear emitting regions, LINERs, which are likely the
lethargic cousins of the luminous active galaxies.  About one thind of
all galaxies show evidence of low-power AGNs, seen in the form of
either LINERs or low-luminosity Seyferts \markcite{Ho:99}(\eg Ho 1999).   This
briefly-described unification model is supported by a variety of
observations, including reverberation mapping, polarization studies,
and mapping of the ionized gas morphology.

Surveys of bright galaxies have characterized the local demographics of
AGN:  the fraction of field galaxies which are classified as Seyferts
rises from approximately $1$\% at $M_{\rm pg} = -21$ to essentially all
galaxies at $M_{\rm pg} = -23$ \markcite{Meurs:82, Meurs:84}(Meurs
1982; Meurs \& Wilson 1984).  The local ratio of obscured (Type~II) to
unobscured (Type~I) Seyferts is approximately 4:1
\markcite{Osterbrock:88}(Osterbrock \& Shaw 1988).  At high redshift,
the Type~I quasars are well-characterized, with the luminosity function
measured out to $z \sim 5$ and individual examples known out to
redshift $z = 6.28$ \markcite{Fan:01}(Fan {et~al.} 2001).  Our
knowledge of Type~II quasars, however, is observationally sparse with
many fewer examples known \markcite{Urry:95}(\eg Urry \& Padovani
1995).  Because of the obscuration, Type~II quasars are not identified
at high redshift from shallow large-area sky surveys, though
low-redshift examples are beginning to be uncovered due to their
unusual colors caused by high-equivalent width emission features
\markcite{Djorgovski:01}(\eg Djorgovski {et~al.} 2001a).  Hard
(2$-$10~kev) X-ray surveys have identified several examples
\markcite{Norman:01}(\eg Norman {et~al.} 2001).  Importantly, but
surprisingly neglected in much of the X-ray Type~II quasar literature,
radio surveys have been detecting the radio-loud end of the Type~II
quasar population for several decades:  these sources are called radio
galaxies \markcite{McCarthy:93}(see McCarthy 1993, for a comprehensive
review) and they have been identified out to redshift $z = 5.19$
\markcite{vanBreugel:99a}(van Breugel {et~al.} 1999).

The abundance of Type~II QSOs at high redshift is of considerable
interest for models of the X-ray background (XRB).  Stellar sources,
hot gas associated with galaxies, and Type~I AGN appear to be
insufficient to explain the XRB.  Most models invoke a substantial
population of luminous, Type~II AGN to produce the 2$-$10~keV
background, with values of the ratio of obscured to unobscured luminous
AGN at $z \sim 1$ ranging from 4:1 to 10:1 \markcite{Madau:94,
Comastri:95}(\eg Madau, Ghisellini, \& Fabian 1994; Comastri {et~al.}
1995).  However, few examples of this population have been observed.
If these sources are indeed shown to exist in such multitudes, they
must indeed make an important contribution to the XRB.

The abundance of Type~II QSOs at high redshift may also prove
interesting in terms of understanding which sources are responsible for
ionizing the early Universe \markcite{Becker:01, Djorgovski:01b}
(Becker {et~al.} 2001; Djorgovski {et~al.} 2001b).  Does the ultraviolet
light produced by hot stars in young protogalaxies ionize the
intergalactic medium at $z \simgt 6$?  Or does ultraviolet light from
obscured and unobscured AGN cause this phase change in the early
Universe?

Several examples of Type~II quasars at low to intermediate redshift
have been suggested from X-ray surveys prior to {\it Chandra} --- \eg
{\it IRAS}~20181$-$2244 at $z=0.185$ \markcite{Elizalde:94}(rest-frame
2$-$10~kev luminosity $L_{2-10} = 2.2 \times 10^{44} \ergs$; Elizalde
\& Steiner 1994), 1E~0449.4$-$1823 at $z=0.338$
\markcite{Stocke:82}($L_{2-10} = 6.7 \times 10^{44} \ergs$; Stocke
{et~al.} 1982), AXJ0341.4$-$4453 at $z=0.672$
\markcite{Boyle:98b}($L_{2-10} = 1.8 \times 10^{44} \ergs$; Boyle
{et~al.} 1988), and RX~J13434+0001 at $z=2.35$ \markcite{Almaini:95,
Georgantopoulos:99}($L_{2-10} \sim 5 \times 10^{45} \ergs$; Almaini
{et~al.} 1995, Georgantopoulos {et~al.} 1999).  In all four cases,
subsequent spectroscopy revealed broad Balmer emission lines, leading
Halpern and collaborators \markcite{Halpern:98a, Halpern:98b,
Halpern:99}(Halpern \& Moran 1998; Halpern, Eracleous, \& Forster 1998;
Halpern, Turner, \& George 1999) to reclassify such Type~II quasar
candidates as narrow-lined Seyfert~1 (NLS1) galaxies -- \ie {\em not}
as high-luminosity analogs of Seyfert~2 galaxies.  According to the
definitions of \markcite{Osterbrock:85}Osterbrock \& Pogge (1985) and
\markcite{Goodrich:89}Goodrich (1989), NLS1s have FWHM H$\beta < 2000
\kms$, \oiii / H$\beta < 3$, and often show emission lines from
\ion{Fe}{2} or higher-ionization iron lines such as
[\ion{Fe}{7}]$\lambda$6087 and [\ion{Fe}{10}]$\lambda$6375.  In
contrast, Seyfert~2 galaxies generally have \oiii / H$\beta > 3$ and no
permitted \ion{Fe}{2} emission.  In terms of studying the XRB the
distinction may be pedantic:  both systems refer to high-luminosity
obscured AGN whose space density at high redshift are poorly known and
thus may be important contributors to the hard (2$-$10~keV) XRB.
However, in terms of studying the physics of active galaxies, the
distinction is far from pedantic.  NLS1 galaxies have visual
absorptions of a few to a few tens of magnitudes while Compton-thick
Seyfert~2 galaxies show visual absorptions of 1000 magnitudes or more,
inferred from their X-ray-determined column densities
\markcite{Halpern:99}(see Halpern {et~al.} 1999).  With distinctly
steeper optical and X-ray spectra and more rapid X-ray variability
relative to normal Seyfert~1s and Seyfert~2s, NLS1s elude explanation
within the orientation-dependent unification scheme of Seyfert
galaxies.  Finally, unlike Seyfert~2 galaxies, NLS1s are not limited to
low luminosity:  several of the PG quasars
\markcite{Boroson:92}(Boroson \& Green 1992) are classified as NLS1s
according to the definitions above.  The lack of high-luminosity
analogs of Seyfert~2s might imply that all sufficiently luminous QSO
nuclei are able to clear Compton-thick absorbing material from their
vicinity, affording detection of the broad-line region from all viewing
angles.

We describe observations of one luminous, high-redshift ($z=3.288$),
Type~II quasar candidate, CXO~J084837.9+445352, hereinafter CXO52,
identified as the 52$^{nd}$ object in our 185~ks {\it Chandra}
observation of the Lynx field \markcite{Stern:01b}(see Stern {et~al.}
2001b).  The Lynx field is one of four fields which comprise the
Spectroscopic Photometric Infrared-Chosen Extragalactic Survey
\markcite{Eisenhardt:01, Stern:01a}(SPICES; Eisenhardt {et~al.} 2001;
Stern {et~al.} 2001a).  CXO52 was identified independently both as an
optical, color-selected, high-redshift source \markcite{Stern:00c}(\eg
Stern {et~al.} 2000c) and from optical follow-up of X-ray sources in
the Lynx field.  The Lynx field contains three known X-ray emitting
clusters out to redshift $z = 1.27$ \markcite{Stanford:97, Rosati:98,
Rosati:99}(Stanford {et~al.} 1997; Rosati {et~al.} 1998, 1999).
Analyses of the {\it Chandra} data for these clusters are described in
\markcite{Stanford:01}Stanford {et~al.} (2001) and
\markcite{Holden:01}Holden {et~al.} (2001).  We note that gravitational
lensing due to the $\approx 30\arcsec$ distant $z=1.27$ galaxy cluster
RX~J0848+4453 has not significantly brightened CXO52.  For the cluster
velocity dispersion $\sigma = 650 \kms$ \markcite{Stanford:01}(Stanford
{et~al.} 2001), CXO52 is magnified by less than 10\%.  In \S2 we
describe the multiwavelength observations of CXO52 which span the X-ray
to radio.  In \S3 we discuss CXO52 in terms of both AGN unification
schemes and the implications for the XRB.  Section~4 summarizes our
results.

We assume $H_0 = 50~ h_{50}~ \kmsMpc$, $\Omega_M = 1$, and
$\Omega_\Lambda = 0$ throughout.  For this cosmology, the luminosity
distance of CXO52 is $26.61~ h_{50}^{-1}$~Gpc and one arcsecond
subtends $7.02~ h_{50}^{-1}$~kpc.   For $H_0 = 65~ \kmsMpc$, $\Omega_M
= 0.35$, and $\Omega_\Lambda = 0.65$, the cosmology favored by recent
high-redshift supernovae and cosmic microwave background observations
\markcite{Riess:01}(\eg Riess {et~al.} 2001), these distances change only slightly:  both
are larger by 9.6\%.

\section{Observations and Results}

\subsection{Ground-Based Optical/Near-Infrared Imaging}

CXO52 was initially identified from deep $BRIz$ imaging using the
``dropout'' color selection techniques which have proved successful at
identifying high-redshift galaxies \markcite{Steidel:96a, Dey:98,
Spinrad:98, Stern:99e}(\eg Steidel {et~al.} 1996; Dey {et~al.} 1998;
Spinrad {et~al.} 1998; Stern \& Spinrad 1999) and quasars
\markcite{Kennefick:95, Fan:99, Stern:00c}(\eg Kennefick, Djorgovski,
\&  de~Calvalho 1995; Fan {et~al.} 1999; Stern {et~al.} 2000c).  The
selection criteria rely upon absorption from the Lyman break and Lyman
forests attenuating the rest-frame ultraviolet continua of
high-redshift sources.  Long-ward of \lya, both star-forming galaxies
and quasars have relatively flat (in $f_\nu$) continua.  Together,
these features provide such systems with a largely unambiguous locus in
multidimensional color-color space.  In concept, the multicolor survey
pursued by a subset of the authors is similar to established quasar
surveys relying upon the digitized Palomar Sky Survey
\markcite{Djorgovski:99}(\eg Djorgovski {et~al.} 1999) and the Sloan
Digital Sky Survey \markcite{Fan:99}(\eg Fan {et~al.} 1999).  In
practice, we probe a much smaller area of sky (eventually a few 100
arcmin$^2$ rather than $\pi - 2 \pi$ steradians) to much fainter
magnitudes (eventually $\approx 27^{th}$ mag AB rather than $\approx
21^{st}$ mag AB).  Our survey has identified numerous sources at $z >
3$, including the faint quasar RD~J030117+002025 at $z = 5.50$ which
was the most distant quasar at the time of its discovery and remains
the lowest luminosity quasar known at $z > 4$
\markcite{Stern:00c}(Stern {et~al.} 2000c).

We obtained optical $BRIz$ imaging using the Kitt Peak National
Observatory (KPNO) 4~m Mayall telescope with its Prime Focus CCD imager
(PFCCD) equipped with a thinned $2048 \times 2048$ array.  This
configuration gives a 16\arcmin $\times$ 16\arcmin\ field of view with
0\farcs47 pix$^{-1}$.  We used a Harris $B$-band ($\lambda_c = 4313$
\AA; $\Delta \lambda = 1069$ \AA), Harris $R$-band ($\lambda_c = 6458$
\AA; $\Delta \lambda = 1472$ \AA), Harris $I$-band ($\lambda_c = 8204$
\AA; $\Delta \lambda = 1821$ \AA), and an RG850 long-pass $z$-band
filter.  The combined, processed images reach limiting AB magnitudes of
26.8 ($B$), 25.6 ($R$), 25.1 ($I$), and 24.7 ($z$), where these numbers
represent 3$\sigma$ limits in 3\arcsec\ diameter apertures.  The
corresponding Vega magnitude limits are 26.9, 25.4, 24.6, and 24.2,
respectively.

We obtained near-infrared $JK_s$ imaging at the KPNO 4~m with the
Infrared Imager \markcite{Fowler:88}(IRIM; Fowler {et~al.} 1988) equipped with a NICMOS~3 256
$\times$ 256 HeCdTe array giving 0\farcs6 pix$^{-1}$.  The $J$-band
($\lambda_c = 1.14 \mu{\rm m}; \Delta \lambda = 0.29 \mu{\rm m}$)
imaging reaches a 3$\sigma$ limiting magnitude of 22.9 (Vega) in a
3\arcsec\ diameter aperture.  The $K_s$-band ($\lambda_c = 2.16 \mu{\rm
m}; \Delta \lambda = 0.33 \mu{\rm m}$) imaging reaches a corresponding
depth of $K_s = 21.4$.  These $BRIzJK_s$ Kitt Peak images comprise one
field in the SPICES field galaxy survey.  Data reduction for the
optical and near-infrared imaging followed standard techniques
\markcite{Eisenhardt:01}(for details, see Eisenhardt {et~al.} 2001).

We supplemented the KPNO optical imaging with deep imaging at Palomar
Observatory.  On UT 1999 November 11$-$12, we used the COSMIC camera
\markcite{Kells:98}(Kells {et~al.} 1998) on the 200\arcsec\ Hale telescope to obtain extremely
deep (5.1~hr) Kron-Cousins $R$-band ($\lambda_c = 6200$ \AA; $\Delta
\lambda = 800$ \AA) imaging of the Lynx field and on UT 2000 April 30
$-$ May 1 we used the same camera/telescope configuration to obtain a
3.2~hr image through a Gunn $i$ filter ($\lambda_c = 8000$ \AA; $\Delta
\lambda = 1800$ \AA).  COSMIC uses a 2048 $\times$ 2048 pixel thinned
CCD with 0\farcs2846 pix$^{-1}$, providing a 9\farcm7 $\times$ 9\farcm7
field of view.  These images, obtained with the purpose of identifying
high-redshift candidates, reach limiting magnitudes of 26.5 and 25.5
mag respectively (Vega; 3$\sigma$ in 3\arcsec\ diameter apertures).
Several $B$- and $R$-band dropouts were identified from the optical
data, many of which have subsequently been confirmed to reside out to
redshifts of $z = 5.63$ and will be discussed in a future manuscript.
Photometry for CXO52 is given in Table~\ref{tablePhot}.  This source
has a relatively flat (in $f_\nu$) spectrum from $R$ to $z$, but shows
a dramatic drop shortward of $R$ with $B - R = 2.2$:  CXO52 is an ideal
$B$-band dropout candidate, exhibiting optical colors corresponding to
a photometric redshift $z \approx 3.5$.  With $I-K = 4.2$, CXO52 also
fits within the conventional definition of an extremely red object
(ERO; $I - K > 4$), an intriguing population likely due to a mixture of
old stellar populations at $z \sim 1.5$ and dusty systems at $z > 1$
\markcite{Dey:99a}(\eg Dey {et~al.} 1999).

% FIGURE 1
\begin{figure}[!t]
\begin{center}
%\plotfiddle{f1a.eps}{1.0in}{0}{45}{45}{-245}{-100}
%\plotfiddle{f1b.eps}{1.0in}{0}{45}{45}{5}{-10}
\end{center}

\caption{Images of CXO52 obtained with WFPC2/F814W ({\bf left}) and
NICMOS/F160W ({\bf right}) on {\it HST}.  The field of view shown is
60\arcsec\ on a side.  North is up, and east is to the left.  CXO52,
located at $\alpha = 08^h48^m37.9^s$, $\delta =
+44\deg53\arcmin51\farcs8$ (J2000), is centered in the images, and is
offset from bright star S (F814W image) by $\Delta \alpha = -11\farcs4,
\Delta \delta = -26\farcs8$.  Contours show {\it Chandra} X-ray imaging
of this field.  The inset, 2\arcsec\ on side, shows the resolved
morphology of CXO52.}

\label{figHST}
\end{figure}                                                                                  

\subsection{{\it HST} Optical/Near-Infrared Imaging}

CXO52 is only 30\arcsec\ ENE of the $z=1.27$ X-ray emitting cluster
RX~J0848+4453 \markcite{Stanford:97, Stanford:01, vanDokkum:01}(Stanford {et~al.} 1997, 2001; van Dokkum {et~al.} 2001) and
therefore lies within {\it Hubble Space Telescope} ({\it HST}) images
of that system obtained with both the Wide Field Planetary Camera~2
\markcite{Trauger:94}(WFPC2; Trauger {et~al.} 1994) and the Near-Infrared Camera and
Multi-Object Spectrograph \markcite{Thompson:99}(NICMOS; Thompson {et~al.} 1999).  The WFPC2
images were taken on UT 1999 March 1 in the F814W ($I_{814}$) filter
and comprise a total of 27.8~ks of integration.  A mosaic of three NIC3
pointings in the F160W ($H_{1.6}$) filter were taken on UT 1998 June
5$-$6 during the campaign when the {\it HST} secondary was moved to
optimized focus for Camera 3.  Each pointing received approximately
11.2~ks of integration.  The {\it HST} imaging and reductions are
discussed in greater detail by \markcite{vanDokkum:01}van Dokkum {et~al.} (2001).

The {\it HST} images of CXO52 are shown in Fig.~\ref{figHST}.  The
galaxy is extended by roughly 1\farcs0 at a position angle of approximately
$-25\deg$.  In 3\arcsec\ diameter apertures, CXO52 has AB
magnitudes of $24.8 \pm 0.2$ through the F814W filter and $23.8 \pm
0.3$ through the F160W filter.  Morphologically, CXO52 appears to be an
interacting system, with an extended southern component, dominating the
measured position angle, and a compact northern component.  Comparison of the
insets in Fig.\ref{figHST} shows a slight chromatic variation in the
morphology of CXO52.  Relative to the F160W image, the centroid of the
northern component appears $\approx 0\farcs1$ farther to the east in
the F814W image.

\subsection{{\it Chandra\/} X-ray Data}

The Advanced CCD Imaging Spectrometer (ACIS-I) on board the {\it
Chandra X-ray Observatory\/} \markcite{Weisskopf:96}(Weisskopf,
{O'dell}, \&  {van~Speybroeck} 1996) obtained a 184.7 ks image of the
Lynx field on UT 2000 May 3$-$4.  \markcite{Stern:01b}Stern {et~al.}
(2001b) discuss reduction of the X-ray data and the identification of
optical and near-infrared counterparts to the 153 compact X-ray sources
in the field.  A total of $\sim 54$ net counts were detected at the
position of the $B$-band dropout galaxy described above.  The source
hardness ratio based on the counts detected in soft (0.5$-$2.0 keV;
$S$) and hard (2.0$-$7.0 keV; $H$) energy bands is $(H-S)/(H+S) = 0.07
\pm 0.13$.  We detect no variability in the X-ray flux or spectrum.
Assuming the flux spectrum can be expressed as a simple power-law
function of energy $E$ (\ie $F(E) \propto E^{\alpha}$) modified by the
Galactic absorption column density in the direction of the source
($N_{\rm H} = 2 \times 10^{20}$ cm$^{-2}$), this hardness ratio
corresponds to a spectral index of $\alpha \approx +0.5$
\markcite{Stern:01b}(Stern {et~al.} 2001b) --- quite unlike the steep
$\alpha \approx -0.8$ spectra that are commonly observed for unabsorbed
AGNs \markcite{Nandra:94}(\eg Nandra \& Pounds 1994).  Given the
narrow-line optical/UV spectrum of the source (see \S 2.5 below), it is
likely that CXO52 actually has an intrinsically steep X-ray spectrum
that appears to be flat due to significant soft X-ray absorption by
intervening material within the galaxy.

To investigate this possibility, we have examined the X-ray spectrum of
CXO52 more closely.  Using version 2.1.2 of the {\it Chandra}
Interactive Analysis of Observations (CIAO) software, we extracted
source counts within a 6\arcsec\ radius aperture centered on the
source; background was measured within a concentric annulus with inner
and outer radii of 10\arcsec\ and 20\arcsec, respectively.  A response
matrix and an effective area file were generated for the location of
the source on chip 0, which is $\sim 3\arcmin$ from the ACIS-I
aimpoint.  The 0.5$-$8~keV spectrum was binned into five channels
containing a minimum of 11 counts each for modeling with XSPEC, the
X-ray spectral fitting package \markcite{Arnaud:96}(Arnaud 1996).
Assuming a Galactic $N_{\rm H} = 2 \times 10^{20}$ cm$^{-2}$, a
power-law fit to the spectrum yields a spectral index of $\alpha =
+0.5$, consistent with the measured hardness ratio.  If we allow for
the possibility of additional absorption at the redshift of the source
and fix the spectral index at a ``typical'' AGN value of $\alpha =
-0.8$, we obtain a best-fit column density of $(4.8 \pm 2.1) \times
10^{23}$ cm$^{-2}$ (90\% confidence) local to CXO52, similar to the
column densities observed in nearby Seyfert~2 galaxies (e.g., Bassani
et al.\ 1999).  In addition, our model implies an unabsorbed 2$-$10 keV
rest-frame luminosity of $3.3 \times 10^{44}$ ergs s$^{-1}$, which is
well within the quasar regime.  Thus, the {\it Chandra\/} data, despite
their limited statistics, provide compelling evidence that CXO52 is
indeed an obscured, luminous AGN.

% FIGURE 2
\begin{figure}[!t]
\begin{center}
\plotfiddle{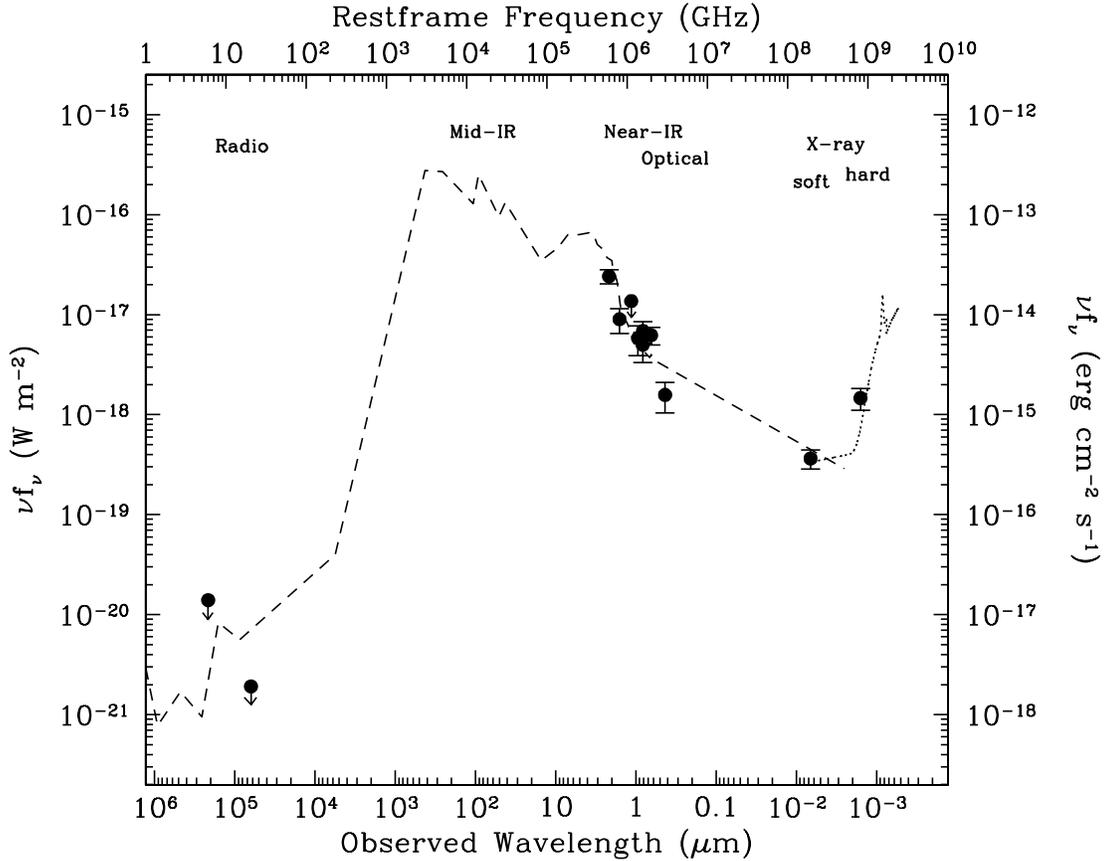}{4.0in}{-90}{60}{60}{-235}{335}
\end{center}

\caption{Photometry for CXO52 plotted with the scaled average spectral
energy distribution (SED) for a Seyfert~2 galaxy.  The dashed line
shows the broad-band (radio to soft X-ray) SED composed from a sample
of 15 local Seyfert~2 galaxies (Schmitt \etal 1997).  The dotted line
shows the composite 1$-$10~keV X-ray spectrum derived by Moran \etal
(2001) from a distance-limited sample of 29 Seyfert 2s observed with
the {\it Advanced Satellite for Cosmology and Astrophysics} ({\it
ASCA}). }

\label{figSED}
\end{figure}

\subsection{Radio Imaging}

CXO52 remains undetected in our deep, 4.8~GHz radio map of the Lynx
field, obtained at the Very Large Array (VLA) on UT 1999 May 29.  The
VLA was in configuration D for these data and the total integration
time spent on source is 11.9~ks.  In source-free regions adjacent to
the quasar, the root-mean-squared (rms) noise of the radio image is
14~$\mu$Jy.  However, deriving an upper limit from the object's
location is complicated by the presence of a 0.29~mJy source $\approx
15\arcsec$ to the west; summing the flux density within a $\approx 330$
arcsec$^2$ synthesized beam area centered on the quasar position yields
a value of 41~$\mu$Jy, dominated by the wings of the adjacent object.
We adopt an upper limit of $f_{\rm 4.8 GHz} = 40~ \mu$Jy.  Comparison
with the 1.4~GHz FIRST (Faint Images of the Radio Sky at
Twenty-Centimeters) survey \markcite{Becker:95}(Becker, White, \& Helfand 1995) reveals no radio source
within 30\arcsec\ of the quasar to a limiting flux density of $f_{\rm
1.4 GHz} \simeq 1$ mJy (5$\sigma$).

Fig.~\ref{figSED} illustrates the broad-band spectral energy
distribution (SED) of CXO52 from our imaging observations which span
nearly 10 orders of magnitude in wavelength.  CXO52 well matches the
average local Seyfert~2 SED derived by \markcite{Schmitt:97}Schmitt
{et~al.} (1997) and \markcite{Moran:01}Moran {et~al.} (2001).  The
observed $B$-band flux is significantly diminished relative to the
composite, naturally explained by absorption due the \lya\ forest for a
source at $z=3.288$.  The hard X-ray spectrum of CXO52 and the
composite Seyfert~2 are indicative of obscured, luminous central
engines.  The non-detection of CXO52 in our radio data suggests that
this galaxy is approximately a factor of three less radio-luminous than
the average Seyfert~2.  \markcite{Ho:01}Ho \& Ulvestad (2001), however,
show that Seyfert nuclei exhibit a wide range in radio properties and
powers.  We conclude that the photometry of CXO52 matches the spectral
shape (SED) of a high-redshift, luminous Seyfert 2, albeit with a
higher bolometric luminosity.

% FIGURE 3
\begin{figure}[!t]
\begin{center}
\plotfiddle{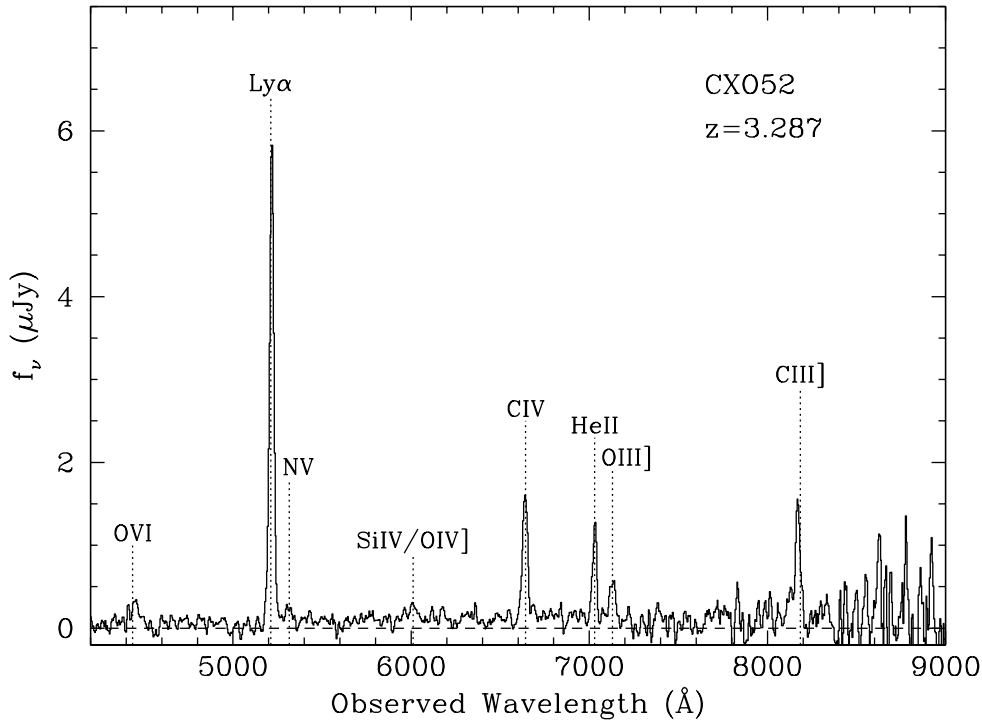}{3.3in}{-90}{50}{50}{-210}{285}
\end{center}

\caption{Optical spectrum of CXO52 obtained with the Keck~I telescope.
The optical spectrum was extracted using a 1\farcs5 $\times$ 1\farcs5
aperture and smoothed with a 15 \AA\ boxcar filter.  Prominent emission
features are labeled.}

\label{figSpec}
\end{figure}

% FIGURE 4
\begin{figure}[!t]
\begin{center}
\plotfiddle{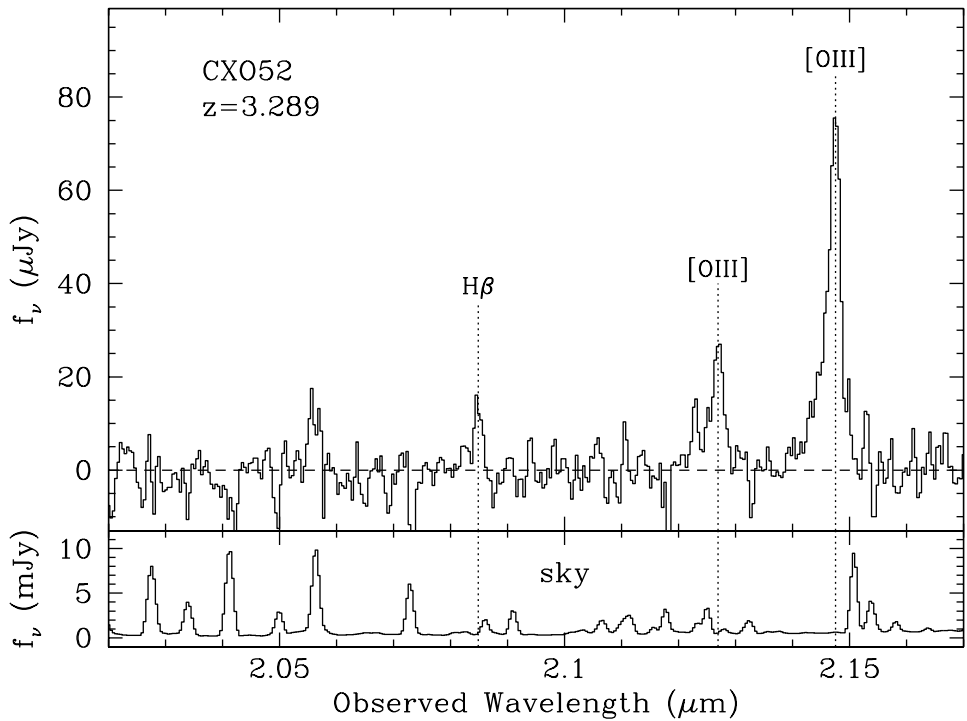}{3.3in}{-90}{50}{50}{-210}{285}
\end{center}

\caption{Near-infrared spectra of CXO52 and the night sky, obtained
with the Keck~II telescope.  The spectra were extracted using 0\farcs57
$\times$ 1\farcs35 apertures.  Prominent emission features are
labeled.  Note that the absorption feature at 2.042 $\mu$m and the
emission feature at 2.056 $\mu$m are both artifacts due to poor
subtraction of telluric OH emission.}

\label{figIRspec}
\end{figure}

\subsection{Optical/Near-Infrared Spectroscopy}

We obtained an optical spectrum of CXO52 on UT 2000 January 11 with the
Low Resolution Imaging Spectrometer \markcite{Oke:95}(LRIS; Oke {et~al.} 1995) on the
Keck~I telescope.  These observations, taken prior to the X-ray
imaging, were in multiobject mode using a slitmask designed to study
high-redshift Lyman-dropout candidates.   The observations, taken at a
position angle of $-108.4\deg$ through a 1\farcs5 slitlet, used the 150
lines mm$^{-1}$ grating ($\lambda_{\rm blaze} = 7500$\AA; resolution $R
\approx 440$) and coarsely sample the wavelength range 4000 \AA\ to
1$\mu$m.  Three equal integrations comprise the 1.5~hr exposure and we
performed 3\arcsec\ spatial offsets between integrations in order to
facilitate removal of fringing at long wavelengths.  As these
observations did not use an order-blocking filter, 2$^{nd}$-order light
might contribute at long wavelengths.  In practice, for a faint, red
source such as CXO52, this contamination is negligible.  We calculated
the dispersion using a NeAr lamp spectrum obtained through the mask
immediately subsequent to the science observations and adjusted the
wavelength zeropoint using telluric emission lines.  The night was
photometric with 0\farcs6 seeing and the spectrum was flux-calibrated
using observations of standard stars from \markcite{Massey:90}Massey \& Gronwall (1990).  The
optical spectrum is shown in Fig.~\ref{figSpec}.

On UT 2001 February 3 we observed CXO52 with the Near Infrared Echelle
Spectrograph \markcite{McLean:98}(NIRSPEC; McLean {et~al.} 1998) at the Nasmyth focus of the
Keck~II telescope.  These observations were taken at a position angle
of $-24.22\deg$ through the 0\farcs57 wide slit in the low dispersion
mode (75 lines mm$^{-1}$ grating), providing 0\farcs193 pix$^{-1}$ and
a resolution $R \approx 1655$.  Our spectrum represents 50 minutes of
integration split into 10 minute dithered exposures, and samples the
wavelength range $1.91 - 2.32 \mu$m.  Wavelength calibration was
performed using a NeAr lamp spectrum obtained through the slit
immediately subsequent to the science observations.  Comparison to
telluric OH emission in those observations show the wavelength solution
to be good to better than 1 \AA.  The spectrum was flux-calibrated
using observations of the B8~V star HD40724 observed at a similar
airmass earlier in the night.  The near-infrared spectrum is shown in
Fig.~\ref{figIRspec}.

The spectra of CXO52 are dominated by several narrow (FWHM $\simgt 1000
\kms$), high-equivalent-width ($W_{\rm \lambda,obs} \simgt 500$ \AA)
emission features.  The unambiguous line identifications and
parameters, calculated for single Gaussian fits using the {\tt SPECFIT}
contributed package within IRAF \markcite{Kriss:94}(Kriss 1994), are presented in
Table~\ref{tableSpec}.  The measured redshift is $3.288 \pm 0.001$.

\section{Discussion}

\subsection{Morphology}

CXO52 is clearly resolved and elongated in the {\it HST} images
(Fig.~\ref{figHST}).  At a redshift of 3.288, the emission extends over
$\approx 7~ h_{50}^{-1}$~kpc.  The WFPC2 F814W filter samples the
rest-frame wavelength range $\lambda \lambda 1700-2200$ \AA\ for CXO52,
which includes the \ciii\ emission line.  The observed equivalent width
of this feature (Table~\ref{tableSpec}) implies that the observed F814W
morphology is strongly dominated by line emission.  The NICMOS F160W
filter samples rest-frame $\lambda \lambda 3300-4200$ \AA\ at this
redshift, which includes the \oiipair\ emission doublet.  No $H$-band
spectrum of CXO52 has been obtained as yet, but based on the current
spectroscopy and typical radio galaxy spectra, the observed F160W morphology is
again likely to be strongly dominated by line emission.  Similarly,
based on the NIRSPEC spectroscopic results, the $K$-band photometry is
dominated by the \oiii\ emission doublet.  Extended emission-line
nebulae are commonly seen in powerful radio galaxies with
spatial scales up to $\approx 400$~kpc \markcite{McCarthy:95}(\eg McCarthy, Spinrad, \& van  Breugel 1995).

The morphologies measured in the {\it HST} images reflect the
distribution of UV light in CXO52, corresponding to a wavelength where
extinction due to dust is most effective.  The peculiar morphology may
be due to a merger system or tidal interactions, or may simply reflect
the distribution of dust in the galaxy.  In particular, the
morphological difference in the location of the northern component
between the F814W and F160W images may indicate a spatially non-uniform
dust screen preferentially reddening one portion of the system.
Similar morphological differences are commonly seen in UV/optical
images of ultraluminous infrared galaxies (ULIGs).

% FIGURE 5
\begin{figure}[!t]
\begin{center}
\plotfiddle{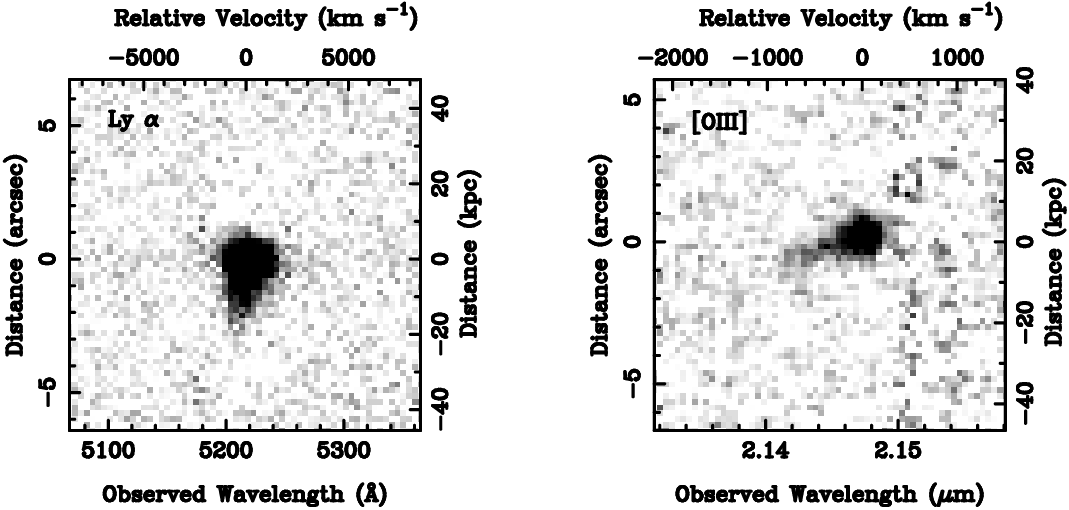}{3.0in}{0}{80}{80}{-245}{-10}
%\plotfiddle{f5a.eps}{1.0in}{0}{45}{45}{-245}{-100}
%\plotfiddle{f5b.eps}{1.0in}{0}{45}{45}{5}{-10}
\end{center}

\caption{Spatially-resolved spectroscopy for the \lya\ and \oiii\
emission lines of CXO52, obtained with the Keck telescopes.  The LRIS
observation of \lya\ ({\bf left}) is through a 1\farcs5 slit at a
position angle of $-108.4$\deg.  The NIRSPEC observation of
\oiii\ ({\bf right}) is through a 0\farcs57 slit at a position able of
$-24.2$\deg.  Velocities are relative to line center.  Distances are
relative to peak of line brightness, with right axes in units of
$h_{50}^{-1}$~ kpc.  Note the sky-line residual at 2.151 $\mu$m in the
infrared spectrum.}

\label{figResolved}
\end{figure}                                                                                  

\subsection{Spatially-Resolved Spectroscopy}

Both \lya\ and \oiii\ are spatially-extended in our two-dimensional
spectra (Fig.~\ref{figResolved}).  The \lya\ profile shows a triangular,
or delta, structure, typical of high-redshift \lya\ emission:
foreground neutral hydrogen from large scale outflows absorbs the blue
wing of the emission line and leads to a P-Cygni profile
\markcite{Dey:98}(\eg Dey {et~al.} 1998).  The \lya\ line of CXO52 is extended by
3\arcsec, or 20~$h_{50}^{-1}$~kpc.  Extended \lya\ emission is commonly
seen in radio galaxies and radio-loud Type-I quasars, though not as
commonly in radio-quiet Type-I quasars.

The \oiii\ emission line of CXO52 has a tadpole morphology, comprised
of a head at a redshift corresponding to the other emission lines of
this source, and a tail extending by $\approx$ 0\farcs5
(3~$h_{50}^{-1}$~kpc) and $\approx 500 \kms$.  The position angle for
these infrared spectra roughly align with the extended morphology seen
in the {\it HST} images.  We may be seeing rotation from CXO52, or the
apparent kinematics may be due to infall/outflow.  Assuming the system
is in dynamic equilibrium, the mass implied by this apparent rotation
is $\approx 2 \times 10^{10} M_\odot$.

\subsection{CXO52 Compared to Other {\it Chandra}-Selected Type~II Quasars}

In both of the mega-second {\it Chandra} deep survey fields, at least
one example of a $z > 2$ Type~II quasar has been identified:
\markcite{Dawson:01}Dawson {et~al.} (2001) identify
CXOHDFN~J123635.6+621424 at $z = 2.011$ in the {\it Hubble} Deep Field
North (HDF-N) and \markcite{Norman:01}Norman {et~al.} (2001) identify
CDF-S~202 at $z = 3.700$ in the {\it Chandra} Deep Field South
(CDF-S).  Many of the properties of CXO52 are similar to these two
sources:  optical spectra of all three show narrow, high-ionization
lines and all three show flat X-ray spectra, indicative of absorption.
For CDF-S~202, \markcite{Norman:01}Norman {et~al.} (2001) infer a local
hydrogen column with $N_H \sim 10^{24} - 10^{25} {\rm cm}^{-2}$ at the
quasar redshift, based on the X-ray spectrum.  The implied rest-frame,
hard-band X-ray luminosity is then $L_{2-10} \sim 10^{45 \pm 0.5}
\ergs$, slightly more luminous than CXO52.  Both hard X-ray
luminosities are comparable to those observed by {\it ASCA} for
optically-selected Type~I quasars \markcite{George:00}(\eg George
{et~al.} 2000).  The optical/near-infrared identification of the
Type~II quasar in the HDF-N has a very red color, $I - K = 4.14$,
similar to that of CXO52 ($I - K_s = 4.21$).  Curiously, however,
CDF-S~202 is quite blue with $I - K = 1.66$, perhaps due to \civ\ and
\heii\ contributions to the observed $I$-band flux.

Neither CXO52 nor CDF-S~202 are detected at radio wavelengths, while
CXOHDFN~J123635.6+621424 is detected in the ultradeep 1.4~GHz VLA
survey of the HDF-N at a flux density of $f_{\rm 1.4 GHz} = 87.7~
\mu$Jy \markcite{Richards:00}(Richards 2000).  For a typical observed
radio galaxy spectral index $\alpha = -1$, where $f_\nu \propto
\nu^\alpha$, the implied rest-frame 1.4~GHz radio luminosity density of
CXOHDFN~J123635.6+621424 is $L_{\rm 1.4 GHz} = 2.5 \times 10^{31}~
h_{50}^{-2}~ \ergsHz$, while $L_{\rm 1.4 GHz} < 1.1 \times 10^{32}~
h_{50}^{-2}~ \ergsHz$ for CDF-S~202 and $L_{\rm 1.4 GHz} < 1.2 \times
10^{32}~ h_{50}^{-2}~ \ergsHz$ for CXO52.  For $\alpha = -0.5$, a
typical radio spectral index for radio-loud quasars, these luminosity
densities drop by a factor of $\approx 3$.  Adopting the
\markcite{Gregg:96}Gregg {et~al.} (1996) cutoff value of $L_{\rm 1.4
GHz} = 10^{32.5} h_{50}^{-2} \ergsHz$ to separate radio-loud sources
from radio-quiet sources, all three of these high-redshift {\it
Chandra}-identified Type~II quasars are radio-quiet.  We note that more
recent work from the FIRST Bright Quasar Survey
\markcite{White:00}(FBQS; White {et~al.} 2000) suggests that no strong
bimodality exists between radio-loud and radio-quiet Type-I quasars.

There is one other likely Type~II quasar in the deep {\it Chandra}
fields:  the mJy radio source VLA~J123642+621331 has recently been
detected in the soft X-ray band by {\it Chandra}
\markcite{Brandt:01}(Brandt {et~al.} 2001).  VLA~J123642+621331, which
lies just outside the HDF-N, is a $f_{\rm 1.4 GHz} = 470\mu$Jy radio
source \markcite{Richards:00}(Richards 2000), shown by
\markcite{Waddington:99}Waddington {et~al.} (1999) to reside at a
likely redshift $z = 4.424$.  The optical spectrum is similar to a
faint radio galaxy or Type~II AGN (${\rm FWHM}_{\rm Ly\alpha} = 420 \pm
75 \kms$).  The derived {\it unabsorbed} rest-frame X-ray luminosity of
$L_{0.5-2} = 2.5 \times 10^{42} h_{50}^{-2} \ergs$ is below the knee of
the \markcite{Miyaji:00}Miyaji, Hasinger, \& M.Schmidt (2000) soft
X-ray luminosity function while the rest-frame 1.4~GHz specific
luminosity, $L_{\rm 1.4 GHz} = 7.1 \times 10^{32}~ h_{50}^{-2}
\ergsHz$, is classified as radio-loud according to the
\markcite{Gregg:96}Gregg {et~al.} (1996) definition.  The likelihood of
finding such a radio-luminous, distant galaxy within the small area of
a single {\it Chandra} pointing is small (see \S 3.6).

\subsection{CXO52 Compared to High-Redshift Radio Galaxies}

With high-ionization emission lines with widths $\simgt 1000 \kms$ and
a spatially-resolved morphology, CXO52 has very similar properties to
high-redshift radio galaxies \markcite{McCarthy:93}(HzRGs; \eg McCarthy
1993), the only flavor of obscured (\ie Type-II) AGN which has been
extensively studied beyond a redshift of unity.  As opposed to quasars
which have broad permitted lines and optical morphologies dominated by
an unresolved nucleus, HzRGs have narrow (FWHM $< 2000 \kms$) permitted
lines and host galaxies which are spatially extended.  Equivalent
widths of forbidden lines are larger in HzRGs than in quasars.

\heii\ is strongly detected in CXO52 with \civ/\heii\ $\approx 2$.
This is typical of HzRGs and atypical of Type~I quasars:  composite
HzRGs spectra show \civ/\heii\ $\simeq 1.5$ \markcite{McCarthy:93,
Stern:99a}(McCarthy 1993; Stern {et~al.} 1999), while composite quasar spectra show \civ/\heii\ $\approx 10
- 50$ \markcite{Boyle:90, VandenBerk:01}(Boyle 1990; {Vanden~Berk} {et~al.} 2001).  Assuming an \oii/\oiii~ ratio
similar to HzRGs, the \oii\ emission line luminosity of CXO52 is
$\approx 5 \times 10^{42} \ergs$, typical of bright 3CR HzRGs
\markcite{McCarthy:93, Willott:99}(McCarthy 1993; Willott {et~al.} 1999).  We reiterate that CXO52 remains {\it
undetected} in our radio images which, at 4.8~GHz, probe {\em four}
orders of magnitude deeper than classical HzRG surveys such as the
MIT-Greenbank survey \markcite{Bennett:86, Lawrence:86}(Bennett {et~al.} 1986; Lawrence {et~al.} 1986) and an order of
magnitude deeper than more modern HzRG surveys such as the Leiden
Berkeley Deep Survey \markcite{Waddington:01}(LBDS; Waddington {et~al.} 2001).  In many ways,
CXO52 appears to be an example of an oxymoronic {\em radio-quiet radio
galaxy}.  Alternatively, just as we know of radio-loud and radio-quiet
(Type-I) quasars, HzRGs simply represent the radio-loud portion of the
Type-II quasar population while CXO52 represents a rare example of a
radio-quiet Type-II quasar.  As discussed in \S 3.5, these radio-quiet
Type-II quasars conceivably outnumber the HzRG population by a factor of a
few:  the rarity of known examples of the population, such as CXO52,
may merely represent observational bias, not an inherent scarcity of the
population.

The \nv\ emission in CXO52 is weak, with \nv / \civ\ $= 0.17$,
approximately half what is seen in composite HzRG spectra
\markcite{McCarthy:93, Stern:99a}(McCarthy 1993; Stern {et~al.} 1999).  \markcite{Vernet:01}Vernet {et~al.} (2001) calculate
photoionization models of HzRGs for a range of metallicity.  The
location of CXO52 in the \ion{N}{5}/\ion{He}{2} versus
\ion{N}{5}/\ion{C}{4} line diagnostic diagram \markcite{Vernet:01}(Fig.~9
of Vernet {et~al.} 2001) implies a metallicity of $\approx 0.8 Z_\odot$.

\markcite{McCarthy:92}McCarthy, Elston, \& Eisenhardt (1992),
\markcite{Eales:93, Eales:96}Eales \& Rawlings (1993, 1996), and
\markcite{Evans:98}Evans (1998) have successfully obtained infrared
spectra of HzRGs from 4~m-class telescopes and show that the line
ratios are most consistent with Seyfert~2s, \ie obscured AGN.
\markcite{Larkin:00}Larkin {et~al.} (2000) reports on NIRSPEC
spectroscopy of the $z=2.630$ HzRG MRC~2025$-$218.  From \lya\ to
\oiii, the spectrum of MRC~2025$-$218 is very similar to that of
CXO52:  both sources show narrow, high-ionization emission and have
\oiii / H$\beta$ line ratios $> 10$, consistent with a Seyfert~2
galaxy.  However, a $K$-band spectrum of MRC~2025$-$218 reveals broad
(FWHM $= 9300 \kms$) H$\alpha$ emission, implying this HzRG is more
like a high-luminosity analog of a Seyfert~1.8 than the high-luminosity
analog of a Seyfert~2.  It is conceivable a longer-wavelength spectrum
of CXO52 would reveal broad H$\alpha$ at $2.8 \mu$m, but such an
observation is beyond current capabilities.  On the other hand,
\markcite{Motohara:01}Motohara {et~al.} (2001) report on near-infrared
spectra of the $z = 2.39$ HzRG LBDS~53W002, obtained with Subaru
telescope.  They detect the rest-frame optical emission lines from
\oii\ to H$\alpha$, finding a Seyfert~2-like \oiii / H$\beta$ line
ratio, and they do not resolve H$\alpha$ (FWHM $\simlt 700 \kms$).

As initially pointed out by \markcite{Lilly:82}Lilly \& Longair (1982),
there exists a surprisingly tight correlation between the 2.2~$\mu$m
($K$ band) infrared magnitude and galaxy redshift for HzRGs, despite
significant morphological evolution \markcite{vanBreugel:98}(van
Breugel {et~al.} 1998) and dramatic $k$-correction effects out to $z =
5.19$, the highest redshift for which a HzRG has been identified
\markcite{vanBreugel:99a}(van Breugel {et~al.} 1999).  At $z = 5.19$,
$K$ samples the rest-frame $U$-band.  With $K = 20.5$, CXO52 is
significantly fainter at 2 $\mu$m than the typical HzRG which has $K
\approx 19$ at $z \approx 3$ \markcite{DeBreuck:01}(\eg {De~Breuck}
{et~al.} 2001).  Does this teach us something about the host galaxies
of radio-quiet versus radio-loud Type-II quasars?  Though the
conventional wisdom that radio-loud Type-I quasars live in ellipticals
and radio-quiet Type-I quasars live in spirals has been debunked
\markcite{McLeod:01}(McLeod \& McLeod 2001, and references
therein), the following correlations appear to remain true:  (1) the
fraction of Type-I quasars inhabiting elliptical host galaxies
increases with nuclear luminosity, and (2) radio-loud Type-I quasars
have brighter host galaxies than radio-quiet Type-I quasars
\markcite{Boyce:98}(\eg Boyce {et~al.} 1998).  Unified models of AGN
remain uncertain as to the cause of these correlations.  One suggestion
is that dust in late-type systems quenches radio jets, preventing their
escape into the intergalactic medium.  In another model,
\markcite{Wilson:95}Wilson \& Colbert (1995) suggest radio-loud AGN are
the products of coalescing supermassive black holes in galaxy mergers,
a process which would result in rapidly spinning black holes capable of
generating highly collimated jets and powerful radio sources.  Such
models would naturally explain the difference in host morphology for
radio-loud and radio-quiet Type-I quasars.  The relative $K$-band
faintness of CXO52 compared to HzRGs at similar redshift may simply
indicate that obscured AGN obey the same correlations.

\subsection{Search Techniques for Type II Quasars}

Because they are much less luminous than Type~I quasars at optical
wavelengths, Type~II quasars suffer from a dearth of observational
study.  In the low-redshift Universe, large sky surveys have begun to
identify these sources due to their unusual colors caused by
high-equivalent width emission features \markcite{Djorgovski:01}(\eg
Djorgovski {et~al.} 2001a).  The radio-loud portion of the
high-redshift Type~II quasar population has been identified through
dedicated, often heroic (\eg Spinrad 1972 $-$ 1998), follow-up of radio
surveys.  However, only $\sim 10 - 15$\%\ of Type~I quasars are
radio-loud, with no evidence of redshift dependence out to $z \simgt 4$
\markcite{Stern:00a}(Stern {et~al.} 2000b).  Are HzRGs just the tip of
the obscured, Type~II quasar iceberg?  Is there a population of
radio-quiet Type~II quasars remaining to be studied with more than six
times the space density of HzRGs?  How else might obscured quasars be
identified, and what can we say about the density of high-redshift
Type~II quasars from the current samples?

As seen from the examples of CXO52 and the Type~II quasars in the {\it
Chandra} mega-second fields, obscured quasars are readily identified
from deep X-ray surveys.  Deep, multi-band imaging surveys are also
sensitive to high-redshift, Type-II quasars:  though they are
significantly fainter than unobscured quasars, Type-II quasars are
readily identified in surveys for distant galaxies.  Indeed, as
mentioned in \S2.1, CXO52 was initially identified in this manner,
prior to {\it Chandra} imaging of the Lynx field.  We have found a
similar source at $z = 2.82$ in the Pisces field of the SPICES survey
(Bower et. al., in preparation).

Due to their high equivalent width, narrow features, obscured quasars
can also be identified in narrow-band imaging surveys.  Several such
surveys are currently in progress with the goal of identifying
high-redshift \lya\ emission due to its high equivalent width
\markcite{Rhoads:01, Hu:99}(\eg Rhoads {et~al.} 2001; Hu, McMahon, \&
Cowie 1999).  Magnitude-limited surveys rarely find galaxies with
emission lines whose rest-frame equivalent width is larger than
$\approx 100$ \AA, the exceptions being \lya\ and H$\alpha$ which can
have rest-frame equivalent widths of 200 \AA\ and 2000 \AA,
respectively, in a young, star-forming galaxy \markcite{Charlot:93}(\eg
Charlot \& Fall 1993).  H$\alpha$ is easily identified from the
neighboring emission lines of \nii\ and
\sii\ \markcite{Stockton:98}(but see Stockton \& Ridgway 1998), as well
as the broad-band SED of the galaxy.  With the $(1 + z)$ boosting of
observed equivalent width due to redshifting, galaxies identified with
$W_\lambda \simgt 200$ \AA\ should almost exclusively be identified
with high-redshift \lya\ \markcite{Stern:00d}(but see Stern {et~al.}
2000a).  However, comparing emission-line-selected samples to
magnitude-limited samples is not quite fair:  it is quite conceivable
that new populations will appear from these emission-line surveys.  In
particular, the known Type~II quasars all show high equivalent width
emission (\eg Fig.~3).  Obscured AGN, interesting in their own right,
are apt to contaminate these high-redshift \lya\ surveys.  Indeed,
\markcite{Rhoads:01}Rhoads {et~al.} (2001), in their Large-Area
Lyman-Alpha (LALA) survey, report an object with narrow \civ, \heii,
and \oiiip\ emission at $z = 2.57$, likely classifiable as a Type~II
AGN.

\subsection{The Density of Type II Quasars and their Contribution to the XRB}

\markcite{Moran:01}Moran {et~al.} (2001) recently derived the local X-ray volume emissivity of
Seyfert~2 galaxies and show that under reasonable assumptions regarding
the spectral energy distribution and evolutionary history of obscured
AGN, they provide an excellent match to the spectrum and intensity of
the cosmic X-ray background (XRB).  Before the XRB problem can be considered
solved, however, these assumptions must be tested.  We next consider
whether the observed numbers of high-redshift Type~II quasars identified in the
deepest {\it Chandra} fields are consistent with expectations and
predictions.

As one estimate of the expected surface density of Type~II quasars, we
can begin with the known distribution of radio-loud AGN.
\markcite{Willott:01}Willott {et~al.} (2001) have recently provided the most accurate
determination of the radio luminosity function (RLF) to date, derived
from several low-frequency surveys.  They fit the RLF with a
dual-population model comprised of:  (1) low-luminosity radio sources,
typically showing weak emission lines; and (2) high-luminosity radio
sources, associated with HzRGs and quasars.  It is this latter
population that is of interest here.  For $2 < z < 5$, Model~C of
\markcite{Willott:01}Willott {et~al.} (2001) reports a surface density of $1.16 \times 10^{-3}$
high-luminosity radio sources per arcminute$^2$, corresponding to 0.33
such sources per $17\arcmin \times 17 \arcmin$ {\it Chandra}/ACIS-I
field.  Approximately 40\%\ of the high-luminosity radio sources are
broad-lined quasars \markcite{Willott:00}(Willott {et~al.} 2000), implying 0.20 narrow-lined
HzRGs per {\it Chandra} field.  \markcite{Stern:00a}Stern {et~al.} (2000b) show that $\approx
12$\% of optically-selected broad-lined quasars are radio-loud, with
little redshift dependence in this fraction out to $z \simgt 4$.  If
Type~II quasars have the same radio-loud fraction, we would expect {\em
one to two} Type~II quasars per {\it Chandra} field at $2 < z < 5$.
Indeed, this is exactly what has been found from preliminary
investigations of the three best-studied, deep {\it Chandra} fields!

We can also estimate the expected surface density of high-redshift
Type~II quasars by scaling from better-constrained Type~I quasar
luminosity functions (QLFs).  Using the same assumptions about
radio-loudness fraction as before, the \markcite{Willott:01}Willott
{et~al.} (2001) high-luminosity radio luminosity function predicts 1.1
unobscured quasars at $2 < z < 5$ per {\it Chandra} field.  The
\markcite{Miyaji:00}Miyaji {et~al.} (2000) Type~I soft X-ray luminosity
function predicts 1.8 unobscured, luminous ($L_{0.5-2} \geq 10^{44.5}
\ergs$) quasars at $2 < z < 5$ per {\it Chandra} field, similar to the
radio prediction.  The Sloan Digital Sky Survey (SDSS) high-redshift,
luminous optical QLF \markcite{Fan:00b}(Fan {et~al.} 2000) predicts
slightly more unobscured quasars per {\it Chandra} field:  it predicts
1.86 such sources with absolute magnitude at 1450 \AA, $M_{1450} < -24$
and 7.92 such sources with $M_{1450} < -23$.  These numbers should be
considered estimates, however, as the SDSS QLF has been derived using a
sample of quasars at higher redshift ($3.6 < z < 5.0$) and brighter
absolute magnitude ($-27.5 < M_{1450} < -25.5$).  We combine the above
results to predict approximately one or two $2 < z < 5$ Type~I quasars
per {\it Chandra} field.  As a baseline we assume a ratio of obscured
(Type~II) to unobscured (Type~I) high-redshift quasars of 3:2, matching
the \markcite{Willott:00}Willott {et~al.} (2000) high-luminosity radio
results, implying one to three $2 < z < 5$ Type~II quasars per {\it
Chandra} field.  If X-ray follow-up of the deep {\it Chandra} fields
finds a substantially larger surface density of high-redshift Type~II
quasars, this would imply that the ratio of obscured to unobscured
luminous AGN at high-redshift was different, with some dependency on
the radio properties of the source.  The implications would be
significant for studies of the XRB and AGN unification scenarios.

\section{Conclusions}

We report the discovery of CXO52, an obscured, or Type~II quasar at $z
= 3.288$, identified independently both as a $B$-band Lyman-dropout
galaxy and as a hard X-ray source in a 185~ks {\it Chandra} exposure.
Besides (1) X-ray selection and (2) color-selection, high-redshift
Type~II quasars are also identifiable from (3) radio surveys and (4)
narrow-band imaging surveys.  This population of misdirected or
obscured quasars, long thought necessary to explain the X-ray
background and demanded by unified models of AGN, are finally being
identified in both their radio-quiet and radio-loud flavors.  They are
likely more populous than the well-studied unobscured, Type~I quasar
population.  {\it Chandra} has now identified examples of radio-quiet,
high-redshift, Type~II quasars in each of deepest fields.  Preliminary
results are consistent with the simple assumption that the ratio of
Type~II systems to Type~I systems is the same for both radio-loud and
radio-quiet systems at high redshift, namely a ratio of 3:2.

We report on the panchromatic properties of CXO52, finding the
broad-band SED is similar to that of local composite Seyfert 2
galaxies.  Morphologically, CXO52 is resolved and somewhat elongated,
distinctly different in appearance from traditional Type~I quasars.
Optical and near-infrared spectra of CXO52 obtained with the Keck
telescopes show strong, narrow emission lines with ratios similar to
HzRGs and Seyfert 2s.

Type~II AGN are luminous in the mid-infrared:  radiation which is
absorbed by gas and dust in the rest-frame ultraviolet/optical escapes
as thermal emission in the mid-infrared.  Several examples of obscured,
reddened quasars have been identified from shallow, wide-area surveys
such as FIRST and 2MASS \markcite{Becker:97, Gregg:01}(\eg Becker
{et~al.} 1997; Gregg {et~al.} 2001).  The {\it Space Infrared Telescope
Facility} ({\it SIRTF}) will identify the unlensed, less extreme
portions of this population.  To first order, luminous AGN are
isotropic at radio, mid- to far-infrared, and hard X-ray wavelengths,
though their ultraviolet/optical properties depend heavily upon
orientation.  Though it is extremely faint optically with $R = 25$,
CXO52 should be luminous in the mid-infrared with $S_{\rm 3.6\mu m}
\approx 50 \mu$Jy and $S_{\rm 24 \mu m} \approx 1.5$~mJy, determined
from the scaled \markcite{Schmitt:97}Schmitt {et~al.} (1997) Seyfert~2
SED (Fig.~2).  These flux levels are easily detectable in a shallow
{\it SIRTF} observation.  Surveys with {\it SIRTF} and the new
generation of X-ray satellites should help provide an unbiased census
of AGN in the Universe, thereby testing models of the XRB and providing
a history of accretion-driven energy production in the Universe.

\acknowledgments

We gratefully acknowledge Carlos De~Breuck for carefully reading the
manuscript and providing insightful comments.   We also thank Aaron
Barth for helpful suggestions.  The authors wish to extend special
thanks to those of Hawaiian ancestry on whose sacred mountain we are
privileged to be guests.  Without their generous hospitality, many of
the observations presented herein would not have been possible.  We
thank Andrea Gilbert for assisting with the NIRSPEC spectroscopy.  Some
of the LRIS data were obtained in the course of a collaborative project
with S.G. Djorgovski.  We thank Niruj Mohan for providing the 4.8~GHz
VLA image.  Support for this work was provided by the National
Aeronautics and Space Administration through {\it Chandra} Award Number
GO0-1082B issued by the {\it Chandra} X-Ray Observatory Center, which
is operated by the Smithsonian Astrophysical Observatory for and on
behalf of NASA under contract NAS8-39073.  The work of DS and PE were
carried out at the Jet Propulsion Laboratory, California Institute of
Technology, under a contract with NASA.  The work of SD was performed
under the auspices of the U.S. Department of Energy, National Nuclear
Security Administration by the University of California, Lawrence
Livermore National Laboratory under contract No.\ W-7405-Eng-48.
E.C.M. is supported by NAS through {\it Chandra} Fellowship grant
PF8-10004, awarded by the {\it Chandra} X-Ray Center, which is operated
by the Smithsonian Astrophysical Observatory for NASA under contract
NAS~8-39073.  This work has also been supported by the following
grants:  NSF grant AST00-71048 (MD), IGPP-LLNL UCRP grant \#02-AP-015
(SD), NSF CAREER grant AST~9875448 (RE), NASA grant NAG 5-6035 (DJH),
and NSF grant AST~95$-$28536 (HS).

%\bibliographystyle{apj}
%% \bibliography

% TABLE 1
\footnotesize
\begin{deluxetable}{ccccc}
\tablecaption{Photometry of CXO52 }
\tablehead{
\colhead{Observed} &
\colhead{Restframe} &
\colhead{Observed} &
\colhead{Flux Density} &
\colhead{Detector/} \\
\colhead{Bandpass} &
\colhead{Bandpass} &
\colhead{Magnitude} &
\colhead{($\mu$Jy)} &
\colhead{Instrument}}
\startdata
2$-$10~keV & 25.7 keV &\nodata& ($1.806\pm0.444$)$\times10^{-4}$ & {\it Chandra}/ACIS \nl
0.5$-$2~keV&  5.4 keV &\nodata& ($1.882\pm0.405$)$\times10^{-4}$ & {\it Chandra}/ACIS \nl
$B$        & 1010 \AA & $27.18^{+0.43}_{-0.31}$ & $0.053\pm0.018$ & KPNO/PFC \nl
$R$        & 1510 \AA & $24.95^{+0.24}_{-0.20}$ & $0.313\pm0.062$ & KPNO/PFC \nl
F814W\tablenotemark{\dag}      & 1900 \AA & $24.81^{+0.18}_{-0.22}$ & $0.433\pm0.082$ & {\it HST}/WFPC2 \nl
$I$        & 1915 \AA & $24.69^{+0.44}_{-0.31}$ & $0.318\pm0.106$ & KPNO/PFC \nl
$z_{AB}$\tablenotemark{\dag}   & 2190 \AA & $24.83^{+0.43}_{-0.31}$ & $0.425\pm0.140$ & KPNO/PFC \nl
$J$        & 2660 \AA & $>22.79$                & $<1.212$        & KPNO/IRIM \nl
F160W\tablenotemark{\dag}      & 3730 \AA & $23.78\pm0.30$          & $1.117\pm0.313$ & {\it HST}/NICMOS \nl
$K_s$      & 5040 \AA & $20.48^{+0.19}_{-0.16}$ & $4.066\pm0.650$ & KPNO/IRIM \nl
3.6~$\mu$m\tablenotemark{\ddag}  & 8400 \AA &\nodata&   45 & {\it SIRTF}/IRAC \nl
4.5~$\mu$m\tablenotemark{\ddag}  & 1.1~$\mu$m&\nodata&   70 & {\it SIRTF}/IRAC \nl
5.8~$\mu$m\tablenotemark{\ddag}  & 1.4~$\mu$m&\nodata&  130 & {\it SIRTF}/IRAC \nl
8.0~$\mu$m\tablenotemark{\ddag}  & 1.9~$\mu$m&\nodata&  270 & {\it SIRTF}/IRAC \nl
24~$\mu$m\tablenotemark{\ddag}   & 5.6~$\mu$m&\nodata& 1500 & {\it SIRTF}/MIPS \nl
4.8~GHz    & 20.6 GHz &\nodata& $<  40$ & VLA \nl
1.4~GHz    &  6.0 GHz &\nodata& $<1000$ & VLA \nl
\enddata

\tablenotetext{\dag}{Observed magnitude in the AB system.}
\tablenotetext{\ddag}{Predicted, assuming local Schmitt \etal (1997)
Seyfert~2 SED.}

\tablecomments{Optical/near-infrared photometry is determined for
3\arcsec\ diameter apertures.  Unless otherwise noted,
optical/near-infrared magnitudes are in the Vega system.  CXO52 remains
undetected in our $J$-band image; we list the 3$\sigma$ limit for the
same aperture instead.}

\label{tablePhot}
\end{deluxetable}
\normalsize

% TABLE 2
\footnotesize
\begin{deluxetable}{lccccc}
\tablecaption{Emission-Line Measurements of CXO52 }
\tablehead{
\colhead{} &
\colhead{$\lambda_{\rm obs}$} &
\colhead{} &
\colhead{Flux} &
\colhead{FWHM} &
\colhead{$W_{\rm \lambda, rest}$} \\
\colhead{Line} &
\colhead{(\AA)} &
\colhead{Redshift} &
\colhead{($10^{-17}$ \ergcm2s)} &
\colhead{(\kms)} &
\colhead{(\AA)}}
\startdata
\ovi       & 4450:             & 3.300: &  1.5:          & 2640:          &  120:         \nl
\lya       & 5217.3  $\pm$ 0.3 & 3.291  & 18.9 $\pm$ 0.4 & 1520 $\pm$  30 & 2100 $\pm$ 40 \nl
\nv        & 5311:             & 3.283: &  0.6:          & 1820:          &   50:         \nl
\sioiv     & 6009:             & 3.283: &  0.4:          & 1320:          &   30:         \nl
\civ       & 6639.2  $\pm$ 0.9 & 3.285  &  3.5 $\pm$ 0.2 & 1350 $\pm$  90 &  350 $\pm$ 20 \nl
\heii      & 7030.4  $\pm$ 1.5 & 3.287  &  1.7 $\pm$ 0.2 &  940 $\pm$ 140 &  170 $\pm$ 20 \nl
\oiiip     & 7130.0  $\pm$ 4.2 & 3.287  &  0.9 $\pm$ 0.3 & 1290 $\pm$ 330 &   90 $\pm$ 30 \nl
\ciii      & 8170.5  $\pm$ 1.4 & 3.280  &  2.1 $\pm$ 0.2 & 1090 $\pm$ 140 &  420 $\pm$ 40 \nl
H$\beta$   & 20847.2 $\pm$ 4.2 & 3.289  &  1.2 $\pm$ 0.8 &  170 $\pm$ 130 &   20 $\pm$ 15 \nl
\oiiia     & 21269.3 $\pm$ 3.4 & 3.289  &  3.7 $\pm$ 1.2 &  300 $\pm$  71 &   50 $\pm$ 20 \nl
\oiii      & 21472.5 $\pm$ 0.7 & 3.288  & 14.4 $\pm$ 0.7 &  430 $\pm$  30 &  380 $\pm$ 20 \nl
\enddata

\tablecomments{All measurements are based on single Gaussian fits to
the emission lines assuming a flat (in $f_\lambda$) continuum.  Line
velocity widths have been deconvolved by the instrument resolution.
Parameters with colons indicate uncertain measurements.  Rest-frame
equivalent widths $W_{\rm \lambda, rest}$ assume $z = 3.288$.}

\label{tableSpec}
\end{deluxetable}
\normalsize


\begin{thebibliography}{}

\bibitem[Almaini, Boyle, Griffiths, Shanks, Stewart, \&  Georgantopoulos 1995]{Almaini:95}
Almaini, O., Boyle, B.~J., Griffiths, R.~E., Shanks, T., Stewart, G.~C., \&  Georgantopoulos, I. 1995, \mnras, 277, L31

\bibitem[Antonucci 1993]{Antonucci:93}
Antonucci, R. 1993, \araa, 31, 473

\bibitem[Arnaud 1996]{Arnaud:96}
Arnaud, K.~A. 1996, in {\it Astronomical Data Analysis Software and Systems V},  ed. J.~G. Jacoby \& J.~Barnes, Vol. 101 (San Francisco: ASP Conference  Series), 17

\bibitem[Becker, Gregg, Hook, McMahon, White, \&  Helfand 1997]{Becker:97}
Becker, R.~H., Gregg, M.~D., Hook, I.~M., McMahon, R.~G., White, R.~L., \&  Helfand, D.~J. 1997, \apjl, 479, 93

\bibitem[Becker, White, \& Helfand 1995]{Becker:95}
Becker, R.~H., White, R.~L., \& Helfand, D.~J. 1995, \apj, 450, 559

\bibitem[Becker {et~al.} 2001]{Becker:01}
Becker, R.~H. {et~al.} 2001, \aj, submitted, astro-ph/0108097

\bibitem[Bennett, Lawrence, Burke, Hewitt, \&  Mahoney 1986]{Bennett:86}
Bennett, C.~L., Lawrence, C.~R., Burke, B.~F., Hewitt, J.~N., \& Mahoney, J.  1986, \apjs, 61, 1

\bibitem[Boroson \& Green 1992]{Boroson:92}
Boroson, T.~A. \& Green, R.~F. 1992, \apjs, 80, 109

\bibitem[Boyce, Disney, Blades, Boksenberg, Crane,  Deharveng, Macchetto, Mackay, \& Sparks 1998]{Boyce:98}
Boyce, P.~J., Disney, M.~J., Blades, J.~C., Boksenberg, A., Crane, P.,  Deharveng, J.~M., Macchetto, F.~D., Mackay, C.~D., {et al.}, 1998,  \mnras, 298, 121

\bibitem[Boyle 1990]{Boyle:90}
Boyle, B.~J. 1990, \mnras, 243, 231

\bibitem[Boyle, Almaini, Georgantopoulos, Blair, Stewart,  Griffiths, Shanks, \& Gunn 1998]{Boyle:98b}
Boyle, B.~J., Almaini, O., Georgantopoulos, I., Blair, A.~J., Stewart, G.~C.,  Griffiths, R.~E., Shanks, T., \& Gunn, K.~F. 1998, \mnras, 297, L53

\bibitem[Brandt {et~al.} 2001]{Brandt:01}
Brandt, W.~N. {et~al.} 2001, \aj, 122, 1

\bibitem[Charlot \& Fall 1993]{Charlot:93}
Charlot, S. \& Fall, S.~M. 1993, \apj, 378, 471

\bibitem[Comastri, Setti, Zamorani, \&  Hasinger 1995]{Comastri:95}
Comastri, A., Setti, G., Zamorani, G., \& Hasinger, G. 1995, \aap, 296, 1

\bibitem[Dawson, Stern, Bunker, Spinrad, \&  Dey 2001]{Dawson:01}
Dawson, S., Stern, D., Bunker, A.~J., Spinrad, H., \& Dey, A. 2001, \aj,  accepted [astro-ph/0105043]

\bibitem[{De~Breuck}, {van~Breugel}, Stanford,  R\"ottgering, Miley, \& Stern 2001]{DeBreuck:01}
{De~Breuck}, C., {van~Breugel}, W., Stanford, S.~A., R\"ottgering, H., Miley,  G., \& Stern, D. 2001, \aj, submitted

\bibitem[Dey, Graham, Ivison, Smail, Wright, \&  Liu 1999]{Dey:99a}
Dey, A., Graham, J.~R., Ivison, R.~J., Smail, I., Wright, G.~S., \& Liu, M.~C.  1999, \apj, 519, 610

\bibitem[Dey, Spinrad, Stern, Graham, \& Chaffee 1998]{Dey:98}
Dey, A., Spinrad, H., Stern, D., Graham, J.~R., \& Chaffee, F. 1998, \apj, 498,  L93

\bibitem[Djorgovski, Castro, Stern, \&  Mahabal 2001a]{Djorgovski:01b}
Djorgovski, S.~G., Castro, S.~M., Stern, D., \& Mahabal, A.~A.  2001a, \apj, submitted, astro-ph/0108069

\bibitem[Djorgovski, Gal, Odewahn, de~Calvalho,  Brunner, Longo, \& Scaramella 1999]{Djorgovski:99}
Djorgovski, S.~G., Gal, R.~R., Odewahn, S.~C., de~Calvalho, R.~R., Brunner, R.,  Longo, G., \& Scaramella, R. 1999, in {\it Wide Field Surveys in Cosmology},  ed. Y.~Mellier \& S.~Colombi (Gif sur Yvette: Editions Fronti\`eres), 89

\bibitem[Djorgovski, Mahabal, Brunner,  Gal, Castro, {de~Calvalho}, \& Odewahn 2001b]{Djorgovski:01}
Djorgovski, S.~G., Mahabal, A.~A., Brunner, R.~J., Gal, R.~R., Castro, S.,  {de~Calvalho}, R.~R., \& Odewahn, S.~C. 2001b, in {\it Virtual  Observatories of the Future}, ed. R.~J. Brunner, S.~G. Djorgovski, \&  A.~Szalay, Vol. 225 (San Francisco: ASP Conference Series), 52

\bibitem[Eales \& Rawlings 1993]{Eales:93}
Eales, S. \& Rawlings, S. 1993, \apj, 411, 67

\bibitem[Eales \& Rawlings 1996]{Eales:96}
---. 1996, \apj, 460, 68

\bibitem[Eisenhardt, Elston, Stanford, Stern, Wu,  Connolly, \& Spinrad 2001]{Eisenhardt:01}
Eisenhardt, P., Elston, R., Stanford, S.~A., Stern, D., Wu, K.~L., Connolly,  A., \& Spinrad, H. 2001, \aj, in preparation

\bibitem[Elizalde \& Steiner 1994]{Elizalde:94}
Elizalde, F. \& Steiner, J.~E. 1994, \mnras, 268, L47

\bibitem[Evans 1998]{Evans:98}
Evans, A.~S. 1998, \apj, 498, 553

\bibitem[Fan {et~al.} 1999]{Fan:99}
Fan, X. {et~al.} 1999, \aj, 118, 1

\bibitem[Fan {et~al.} 2000]{Fan:00b}
---. 2000, \aj, 120, 1167

\bibitem[Fan {et~al.} 2001]{Fan:01}
---. 2001, \aj, submitted [astro-ph/0108063]

\bibitem[Fowler, Gatley, Stuart, Joyce, \&  Probst 1988]{Fowler:88}
Fowler, A.~M., Gatley, I., Stuart, F., Joyce, R.~R., \& Probst, R.~G. 1988,  SPIE, 972, 107

\bibitem[Georgantopoulos, Almaini, Shanks,  Stewart, Griffiths, Boyle, \& Gunn 1999]{Georgantopoulos:99}
Georgantopoulos, I., Almaini, O., Shanks, T., Stewart, G.~C., Griffiths, R.~E.,  Boyle, B.~J., \& Gunn, K.~F. 1999, \mnras, 305, 125

\bibitem[George, Turner, Yaqoob, Netzer, Laor, Mushotzky, Nandra, \& Takahashi 2000]{George:00}
George, I.~M., Turner, T.~J., Yaqoob, T., Netzer, H., Laor, A., Mushotzky, R.~F., Nandra, K., \& Takahashi, T. 2000, \apj, 531, 52

\bibitem[Goodrich 1989]{Goodrich:89}
Goodrich, R.~W. 1989, \apj, 342, 234

\bibitem[Gregg, Becker, White, Helfand, McMahon, \&  Hook 1996]{Gregg:96}
Gregg, M.~D., Becker, R.~H., White, R.~L., Helfand, D.~J., McMahon, R.~G., \&  Hook, I.~M. 1996, \aj, 112, 407

\bibitem[Gregg, Lacy, White, Glikman, Helfand, Becker, \&  Brotherton 2001]{Gregg:01}
Gregg, M.~D., Lacy, M., White, R.~L., Glikman, E., Helfand, D.~J., Becker,  R.~H., \& Brotherton, M.~S. 2001, \apj, in press [astro-ph/0107441]

\bibitem[Halpern, Eracleous, \& Forster 1998]{Halpern:98b}
Halpern, J.~P., Eracleous, M., \& Forster, K. 1998, \apj, 501, 103

\bibitem[Halpern \& Moran 1998]{Halpern:98a}
Halpern, J.~P. \& Moran, E.~C. 1998, \apj, 494, 194

\bibitem[Halpern, Turner, \& George 1999]{Halpern:99}
Halpern, J.~P., Turner, T.~J., \& George, I.~M. 1999, \mnras, 307, L47

\bibitem[Ho 1999]{Ho:99}
Ho, L.~C. 1999, Adv. in Space Research, 23, 813

\bibitem[Ho \& Ulvestad 2001]{Ho:01}
Ho, L.~C. \& Ulvestad, J.~S. 2001, \apjs, 133, 77

\bibitem[Holden, Stanford, Rosati, Tozzi, Eisenhardt, \&  Spinrad 2001]{Holden:01}
Holden, B., Stanford, S.~A., Rosati, P., Tozzi, P., Eisenhardt, P. R.~M., \&  Spinrad, H. 2001, \aj, in press (August)

\bibitem[Hu, McMahon, \& Cowie 1999]{Hu:99}
Hu, E.~M., McMahon, R.~G., \& Cowie, L.~L. 1999, \apj, 522, 9

\bibitem[Kells, Dressler, Sivaramakrishnan, Carr, Koch,  Epps, Hilyard, \& Pardeilhan 1998]{Kells:98}
Kells, W., Dressler, A., Sivaramakrishnan, A., Carr, D., Koch, E., Epps, H.,  Hilyard, D., \& Pardeilhan, G. 1998, \pasp, 110, 1487

\bibitem[Kennefick, Djorgovski, \&  de~Calvalho 1995]{Kennefick:95}
Kennefick, J.~D., Djorgovski, S.~G., \& de~Calvalho, R.~R. 1995, \aj, 110, 2553

\bibitem[Kriss 1994]{Kriss:94}
Kriss, G. 1994, in {\it Astronomical Data Analysis Software and Systems III },  Vol.~61 (San Francisco: ASP Conference Series), 437

\bibitem[Larkin {et~al.} 2000]{Larkin:00}
Larkin, J.~E. {et~al.} 2000, \apj, 533, L61

\bibitem[Lawrence, Readhead, Moffet, \&  Birkinshaw 1986]{Lawrence:86}
Lawrence, C.~R., Readhead, A.~C.~S., Moffet, A.~T., \& Birkinshaw, M. 1986,  \apjs, 61, 105

\bibitem[Lilly \& Longair 1982]{Lilly:82}
Lilly, S.~J. \& Longair, M.~S. 1982, \mnras, 199, 1053

\bibitem[Madau, Ghisellini, \& Fabian 1994]{Madau:94}
Madau, P., Ghisellini, G., \& Fabian, A.~C. 1994, \mnras, 270, 17

\bibitem[Massey \& Gronwall 1990]{Massey:90}
Massey, P. \& Gronwall, C. 1990, \apj, 358, 344

\bibitem[McCarthy 1993]{McCarthy:93}
McCarthy, P.~J. 1993, \araa, 31, 639

\bibitem[McCarthy, Elston, \& Eisenhardt 1992]{McCarthy:92}
McCarthy, P.~J., Elston, R., \& Eisenhardt, P. 1992, \apj, 387, L29

\bibitem[McCarthy, Spinrad, \& van  Breugel 1995]{McCarthy:95}
McCarthy, P.~J., Spinrad, H., \& van Breugel, W. 1995, \apjs, 99, 27

\bibitem[McLean {et~al.} 1998]{McLean:98}
McLean, I.~S. {et~al.} 1998, SPIE, 3354, 566

\bibitem[McLeod \& McLeod 2001]{McLeod:01}
McLeod, K.~K. \& McLeod, B.~A. 2001, \apj, 546, 782

\bibitem[Meurs 1982]{Meurs:82}
Meurs, E.~J.~A. 1982, {\it The Seyfert galaxy population --- A radio survey;  luminosity functions, related objects} (Univ. Leiden: Ph.D. thesis)

\bibitem[Meurs \& Wilson 1984]{Meurs:84}
Meurs, E.~J.~A. \& Wilson, A.~S. 1984, \aap, 136, 206

\bibitem[Miyaji, Hasinger, \& M.Schmidt 2000]{Miyaji:00}
Miyaji, T., Hasinger, G., \& M.Schmidt. 2000, \aap, 353, 25

\bibitem[Moran, Kay, Davis, Filippenko, \&  Barth 2001]{Moran:01}
Moran, E.~C., Kay, L.~E., Davis, M., Filippenko, A.~V., \& Barth, A.~J. 2001,  \apj, in press, astro-ph/0106519

\bibitem[Motohara {et~al.} 2001]{Motohara:01}
Motohara, K. {et~al.} 2001, \pasj, in press, astro-ph/0104473

\bibitem[Nandra \& Pounds 1994]{Nandra:94}
Nandra, K. \& Pounds, K.~A. 1994, \mnras, 268, 405

\bibitem[Norman {et~al.} 2001]{Norman:01}
Norman, C. {et~al.} 2001, \apj, submitted, astro-ph/0103198

\bibitem[Oke, Cohen, Carr, Cromer, Dingizian, Harris,  Labrecque, Lucinio, Schaal, Epps, \& Miller 1995]{Oke:95}
Oke, J.~B., Cohen, J.~G., Carr, M., Cromer, J., Dingizian, A., Harris, F.~H.,  Labrecque, S., Lucinio, R., {et al.}, 1995, \pasp,  107, 375

\bibitem[Osterbrock \& Pogge 1985]{Osterbrock:85}
Osterbrock, D.~E. \& Pogge, R.~W. 1985, \apj, 297, 166

\bibitem[Osterbrock \& Shaw 1988]{Osterbrock:88}
Osterbrock, D.~E. \& Shaw, R.~A. 1988, \apj, 327, 89

\bibitem[Rhoads, Malhotra, Dey, Stern, Spinrad, \&  Jannuzi 2001]{Rhoads:01}
Rhoads, J.~E., Malhotra, S., Dey, A., Stern, D., Spinrad, H., \& Jannuzi, B.~T.  2001, \apj, 545, L85

\bibitem[Richards 2000]{Richards:00}
Richards, E.~A. 2000, \apj, 533, 611

\bibitem[Riess {et~al.} 2001]{Riess:01}
Riess, A.~G. {et~al.} 2001, \apj, in press; astro-ph/0104455

\bibitem[Rosati, {della~Ceca}, Norman, \&  Giacconi 1998]{Rosati:98}
Rosati, P., {della~Ceca}, R., Norman, C., \& Giacconi, R. 1998, \apj, 492, L21

\bibitem[Rosati, Stanford, Eisenhardt, Elston, Spinrad,  Stern, \& Dey 1999]{Rosati:99}
Rosati, P., Stanford, S.~A., Eisenhardt, P.~R., Elston, R., Spinrad, H., Stern,  D., \& Dey, A. 1999, \aj, 118, 76

\bibitem[Schmitt, Kinney, Calzetti, \&  {Storchi-Bergmann} 1997]{Schmitt:97}
Schmitt, H.~R., Kinney, A.~L., Calzetti, D., \& {Storchi-Bergmann}, T. 1997,  \aj, 114, 592

\bibitem[Spinrad, Stern, Bunker, Dey, Lanzetta, Yahil,  Pascarelle, \& Fern\`andez-Soto 1998]{Spinrad:98}
Spinrad, H., Stern, D., Bunker, A.~J., Dey, A., Lanzetta, K., Yahil, A.,  Pascarelle, S., \& Fern\`andez-Soto, A. 1998, \aj, 116, 2617

\bibitem[Stanford, Elston, Eisenhardt, Spinrad, Stern,  \& Dey 1997]{Stanford:97}
Stanford, S.~A., Elston, R., Eisenhardt, P.~R.~M., Spinrad, H., Stern, D., \&  Dey, A. 1997, \aj, 114, 2232

\bibitem[Stanford, Rosati, Holden, Tozzi, Eisenhardt,  \& Spinrad 2001]{Stanford:01}
Stanford, S.~A., Rosati, P., Holden, B., Tozzi, P., Eisenhardt, P.~R.~M., \&  Spinrad, H. 2001, \apj, 552, 504

\bibitem[Steidel, Giavalisco, Dickinson, \&  Adelberger 1996]{Steidel:96a}
Steidel, C.~S., Giavalisco, M., Dickinson, M., \& Adelberger, K.~L. 1996, \aj,  112, 352

\bibitem[Stern, Bunker, Spinrad, \&  Dey 2000a]{Stern:00d}
Stern, D., Bunker, A.~J., Spinrad, H., \& Dey, A. 2000a, \apj,  537, 73

\bibitem[Stern, Connolly, Eisenhardt,  Elston, Holden, Rosati, Stanford, Spinrad, Tozzi, \& Wu 2001a]{Stern:01a}
Stern, D., Connolly, A., Eisenhardt, P., Elston, R., Holden, B., Rosati, P.,  Stanford, S.~A., Spinrad, H., {et al.}, 2001a, in  {\it Deep Fields 2000 } (Berlin: Springer Verlag), in press  [astro-ph/0012146]

\bibitem[Stern, Dey, Spinrad, Maxfield, Dickinson,  Schlegel, \& Gonz\'alez 1999]{Stern:99a}
Stern, D., Dey, A., Spinrad, H., Maxfield, L.~M., Dickinson, M.~E., Schlegel,  D., \& Gonz\'alez, R.~A. 1999, \aj, 117, 1122

\bibitem[Stern, Djorgovski, Perley,  de~Carvalho, \& Wall 2000b]{Stern:00a}
Stern, D., Djorgovski, S.~G., Perley, R., de~Carvalho, R., \& Wall, J.  2000b, \aj, 132, 1526

\bibitem[Stern \& Spinrad 1999]{Stern:99e}
Stern, D. \& Spinrad, H. 1999, \pasp, 111, 1475

\bibitem[Stern, Spinrad, Eisenhardt, Bunker,  Dawson, Stanford, \& Elston 2000c]{Stern:00c}
Stern, D., Spinrad, H., Eisenhardt, P., Bunker, A.~J., Dawson, S., Stanford,  S.~A., \& Elston, R. 2000c, \apj, 533, L75

\bibitem[Stern, Tozzi, Stanford, Rosati,  Holden, Eisenhardt, Elston, Wu, Connolly, Spinrad, Dawson, Dey, \&  Chaffee 2001b]{Stern:01b}
Stern, D., Tozzi, P., Stanford, S.~A., Rosati, P., Holden, B., Eisenhardt, P.,  Elston, R., Wu, K.~L., {et al.}, 2001b, \aj, submitted

\bibitem[Stocke, Liebert, Maccacaro, Griffiths, \&  Steiner 1982]{Stocke:82}
Stocke, J., Liebert, J., Maccacaro, T., Griffiths, R.~E., \& Steiner, J.~E.  1982, \apj, 252, 69

\bibitem[Stockton \& Ridgway 1998]{Stockton:98}
Stockton, A. \& Ridgway, S.~E. 1998, \aj, 115, 1340

\bibitem[Thompson, Storrie-Lombardi, Weymann, Rieke,  Schneider, Stobie, \& Lytle 1999]{Thompson:99}
Thompson, R., Storrie-Lombardi, L.~J., Weymann, R.~J., Rieke, M.~J., Schneider,  G., Stobie, E., \& Lytle, D. 1999, \aj, 117, 17

\bibitem[Trauger {et~al.} 1994]{Trauger:94}
Trauger, S. {et~al.} 1994, \apj, 435, 3

\bibitem[Urry \& Padovani 1995]{Urry:95}
Urry, C.~M. \& Padovani, P. 1995, \pasp, 107, 803

\bibitem[van Breugel, {De~Breuck}, Stanford, Stern,  R\"ottgering, \& Miley 1999]{vanBreugel:99a}
van Breugel, W., {De~Breuck}, C., Stanford, S.~A., Stern, D., R\"ottgering, H.,  \& Miley, G. 1999, \apjl, 518, L61

\bibitem[van Breugel, Stanford, Spinrad, Stern, \&  Graham 1998]{vanBreugel:98}
van Breugel, W., Stanford, S.~A., Spinrad, H., Stern, D., \& Graham, J.~R.  1998, \apj, 502, 614

\bibitem[van Dokkum, Stanford, Holden, Eisenhardt,  Dickinson, \& Elston 2001]{vanDokkum:01}
van Dokkum, P.~G., Stanford, S.~A., Holden, B.~P., Eisenhardt, P.~R.,  Dickinson, M., \& Elston, R. 2001, \apj, in press

\bibitem[{Vanden~Berk} {et~al.} 2001]{VandenBerk:01}
{Vanden~Berk}, D.~E. {et~al.} 2001, \aj, in press, astro-ph/0105231

\bibitem[Vernet, Fosbury, {Villar-Martin}, Cohen,  Cimatti, {di~Serego~Alighieri}, \& Goodrich 2001]{Vernet:01}
Vernet, J., Fosbury, R.~A.~E., {Villar-Martin}, M., Cohen, M.~H., Cimatti, A.,  {di~Serego~Alighieri}, S., \& Goodrich, R.~W. 2001, \aap, 366, 7

\bibitem[Waddington, Dunlop, Peacock, \&  Windhorst 2001]{Waddington:01}
Waddington, I., Dunlop, J.~S., Peacock, J.~A., \& Windhorst, R.~A. 2001,  \mnras, submitted; astro-ph/0107048

\bibitem[Waddington, Windhorst, Cohen, Partridge,  Spinrad, \& Stern 1999]{Waddington:99}
Waddington, I., Windhorst, R.~A., Cohen, S.~H., Partridge, R.~B., Spinrad, H.,  \& Stern, D. 1999, \apj, 526, 77

\bibitem[Weisskopf, {O'dell}, \&  {van~Speybroeck} 1996]{Weisskopf:96}
Weisskopf, M.~C., {O'dell}, S.~L., \& {van~Speybroeck}, L.~P. 1996, SPIE, 2805,  2

\bibitem[White {et~al.} 2000]{White:00}
White, R.~L. {et~al.} 2000, \apjs, 126, 133

\bibitem[Willott, Rawlings, Blundell, \&  Lacy 1999]{Willott:99}
Willott, C.~J., Rawlings, S., Blundell, K.~M., \& Lacy, M. 1999, \mnras, 309,  1017

\bibitem[Willott, Rawlings, Blundell, \&  Lacy 2000]{Willott:00}
---. 2000, \mnras, 316, 449

\bibitem[Willott, Rawlings, Blundell, Lacy, \&  Eales 2001]{Willott:01}
Willott, C.~J., Rawlings, S., Blundell, K.~M., Lacy, M., \& Eales, S.~A. 2001,  \mnras, 322, 536

\bibitem[Wilson \& Colbert 1995]{Wilson:95}
Wilson, A.~S. \& Colbert, E.~J.~M. 1995, \apj, 438, 62

\end{thebibliography}
\end{document}